Topical Discovery of Web Content

Giancarlo Crocetti

A Thesis

Submitted in Partial Fulfillment of the

Requirements for the Degree of

Master of Science in Data Mining

Department of Mathematical Sciences

Central Connecticut State University

New Britain, Connecticut

October 2012

Thesis Advisor

Dr. Roger Bilisoly

Department of Mathematical Sciences

# DEDICATION

This thesis is dedicated to my loving wife Younhee that supported my studies with moving dedication and patience. You are a true inspiration and a compass that always shows me the way to a safe and happy state of mind.

The achievement of this Master's degree would not have been possible without you.

Thank you.

# ACKNOLEDGEMENTS

I'm particularly grateful to Prof. Bilisoly to be my thesis advisor.

Data Mining represented a clear improvement in my career and opened doors that would have been unavailable to me otherwise. Therefore, my deepest gratitude goes to the entire faculty of CCSU that not only guided me through this course with knowledge and professionalism, but also provided a compassionate support, a quality that makes them truly an outstanding team and I bow to them.

I particularly thank Prof. Daniel Larose for his constant encouragement in any moment of difficulty, Prof. Roger Bilisoly for his deep insight into Text Mining and Prof. Dan Miller that was able to inject his love and understanding of the math behind statistics, even to somebody with a natural "repulsion" to the topic like myself.

I thank all my fellow students especially Benjamin Dickman, Rajiv Sambasivan, Jill Willie, Virendra Rawat, Vince Rogers, Abe Weston, Rick Roundtree, Sairam Tadigadapa, Thomas Wilk, and George DeVarennes from which I've learned a great deal.

I thank my parents that always supported me and constantly stressed the value of education and learning.

# Contents

## Abstract

The use of analytics in the enterprise in general, and in business intelligence (BI) in particular, is going through a dramatic change in which the use of unstructured data and search technologies are playing a pivotal role. Researchers, business analysts, competitive intelligence experts, all are leveraging search technologies in order to analyze and generate insight for specific and relevant questions.

Unstructured data sources, related to research topics, are identified and indexed by search engines that will augment the textual content with annotations and transform it into a form suited for information retrieval. Even though this process is proven successful, it contains a weak link represented by the manual selection of new unstructured data sources to be added to the system over time. These data sources represent a corpus of relevant documents related to a specific research area and are in the form of web pages.

Usually, the task of identifying relevant web pages is outsourced to external companies that provide this manual labor-intensive process on a subscription-based fee that can easily total tens of thousands of dollars per year. Moreover, due to the manual nature of the process, this task is prone to many errors and often results in content of poor quality adding more noise than knowledge to the search index.

To this end, this thesis describes the theory and the implementation of the author's new software tool, the "Web Topical Discovery System" (WTDS), which provides an approach to the automatic discovery and selection of new web pages relevant to specific analytical needs. We will see how it is possible to specify the research context with search keywords related to the area of interest and consider the important problem of removing extraneous data from a web page containing an article in order to reduce, to a minimum, false positives represented by a match on a keyword that is showing up on the latest news box of the same page. The removal of duplicates, the analysis of richness of information contained in the article and lexical diversity are all taken into consideration in order to provide the optimum set of recommendations to the end user or system.

# 1. Introduction to the Web Topical Discovery System

In this research we are concerned with the discovery of newly-created content associated with research topics being inputted by a user in the form of keywords or document samples.

We will argue that a classic crawling approach cannot be applied in this case, since crawling is based on previously known web sources and the discovery of new content is completely dependent on hyperlinks (a new page is only discovered if a page links to it) and independent of topics (a link is followed independently on the content of the page). Moreover, new content is most valuable at the time of creation; therefore the overhead of discovering new content through the use of crawlers, that can be translated in weeks or even months, is unacceptable.

In this work, a new discovery paradigm is presented, that relies on external search engines like Yahoo or Bing for the detection of new content. Among possible content candidates, measures of freshness, similarity, lexical diversity, text entropy, etc., are calculated in order to identify and select the top documents that are returned as recommendations to the calling application in the form of URLs.

The following quote from a Nov. 2002 article written by A. Nelson (http://www.craweblogs.com/commlog/archives/000533.html) hints about the importance of measures like lexical diversity that can be used to identify intrinsic properties of a text.

*"Research demonstrates that audiences consistently judge lexically diverse speakers as more competent, of higher social status, and more persuasive than a low-diversity counterpart. The effect is stronger still the more formal the situation, and it doesn't lie only among spoken communication: lexically diverse written messages demonstrate the same positive effects as well, and often to a greater extent."*

If lexical diversity hints on how competent the author of the article is, text entropy might confirm the richness of the vocabulary used and a useful measure to compare richness in two or more texts. Finally, the recommendations generated by the WTDS can then be indexed by a company's internal search engine and the associated content made available to analytical applications that will process the newly available data for insight, knowledge discovery or competitive intelligence.

## 1.1 Current State of Research

Before embarking ourselves in any real effort on this topic, it is important to understand the current state of research in this area and, eventually, the results achieved in the field of content discovery. To our surprise, even though the subject of automatic web site recommendations and content discovery has been extensively studied from a data mining perspective, all efforts are centered on the following specific use cases:

- ***Product Recommendation***: given the product offerings and user's history the system will recommend products close to the user's taste. Examples of this are Amazon and Netflix.
- ***Content Recommendation***: given the user's browser history the system will recommend content the user might be interested in. Even though this subject might be close to our research topic, the recommendations made by this kind of system are based on local content stored within the web site or within a small conglomerate of sites.

We can consider the following, as products with similar functionality:

1. **Thoora:** a content discovery platform particularly tuned to social media. Even though this product positions itself in the same research topic, we have important differences that we must consider:

    a. Thoora searches a big collection of predefined sites. No matter how large the collection is the results will always be extracted within this boundary, potentially missing important sources of data. The fact that you need to add new sources manually undermines Thoora's main purpose.

    b. The discovery is based on keywords only.

    Even though this is a key feature it is also the weakest one since the user must manually add the context around a word. An example of this is the keyword "Washington": do we mean Washington State, Washington DC, or George Washington?

2. **Darwin Awareness Engine**: a more sophisticated engine that monitors specific topics real time. In this case we have a platform that is able to monitor the World Wide Web and look for specific signals; however the end result does not provide recommendations of possible sources (URLs) to add to an in-house search platform.

An interesting aspect of discovery engines is that, in general, their architecture and algorithmic details are patented and not open to the public. Moreover, searches on academic resources like ACM or IEEE did not produce any relevant findings, a possible indication of the novelty of this topic.

## 1.2 Motivation

The use of search engines within the enterprise has dramatically evolved over the past few years and from a tool that allows users to find documents within complex and disparate repositories, it has now become a key component of knowledge discovery and analytics in departments like competitive intelligence and R&D.

This novel use of search engines is dictating much "more" stringent requirements on data freshness and especially data discovery. As well described in his "Real Time is the Rule" address on Dec. 15[th] 2011 in New York during a data virtualization luncheon, Ted Hills showed that the value of data decreases over time and it is the highest right after its creation as shown in Figure 1. Consequently, we must provide to the search engine references to fresh content or, at minimum, information that is still useful for analytic use.

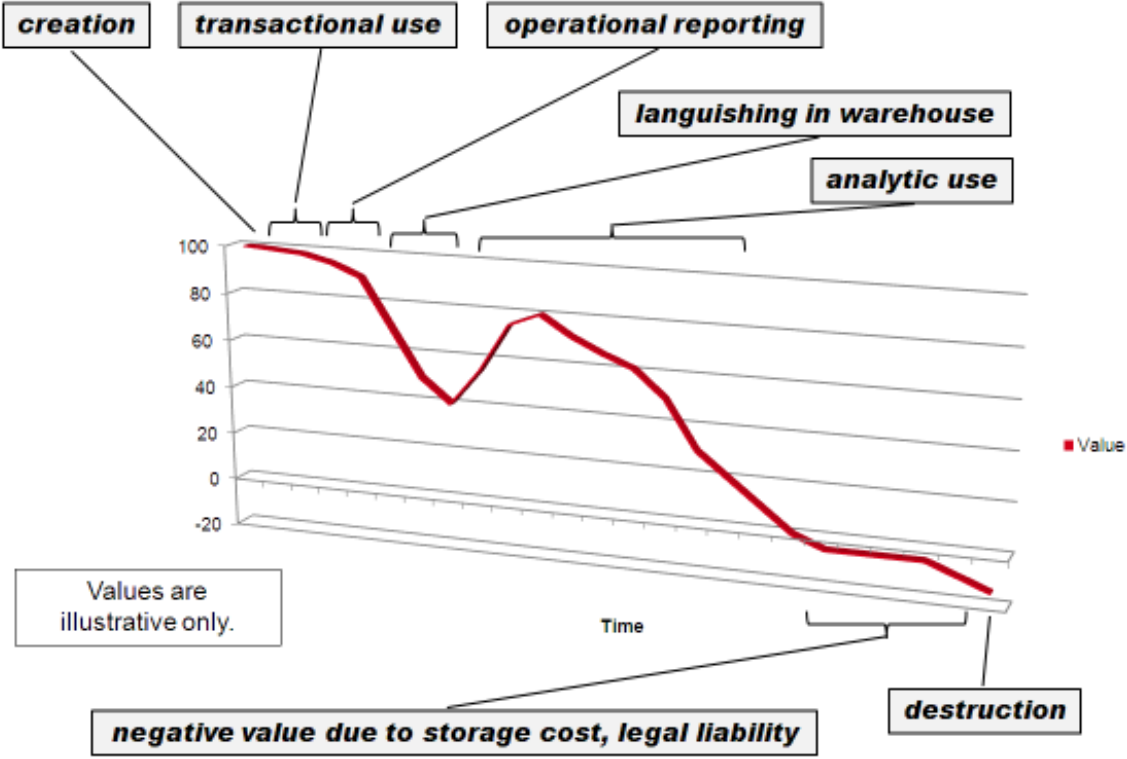

**Fig. 1** – Value of Data Over Time.

## 1.3 A New Paradigm

Search engines rely on web crawlers for the discovery of new content. However, despite the different approaches to content crawling, this technique is not satisfactory for our goal for the following reasons:

1. Crawlers are not topical. Crawlers reach new content exclusively by the use of hyperlinks, they are completely unaware of the content they are crawling. In support to this point in Dec. 2002, D. Sullivan wrote an interesting article titled "Revisiting Meta Tags" in which he states the uselessness of "meta keyword" tags used by search engines during the matching process in a keyword search.
2. Limited Coverage. The content discovered is limited to the link being followed and the crawler is not able to reach out to every page.
3. Inability to correctly estimate importance of URLs. Even though crawlers rank a page based on the analysis of hyper-links, it is not trivial for crawlers to determine the importance of a web page based on its content.

These reasons alone build a strong case for discarding web crawlers as possible tools for content discovery.

A possible alternative comes in the form of worldwide search engines like Google, Yahoo and Bing, to cite the best known, which can be leveraged as sources of content candidates. In fact they possess all the characteristics necessary for our tasks ahead:

1. Full Coverage: being global search engines they contain almost the entirety of public web content. This means that we will not, in general, reach to what is called "Deep Web" since this content is usually not available for indexing in public search engines. In order to increase our reach, we might implement an adapter to submit queries to vertical search engines like Intute.ac.uk.
2. Topical Content: through the use of keywords it is possible to retrieve content specific to particular topics.

3. Freshness: even though each search engine developed its own ranking algorithm, new content tends to rank higher than the existing one.

Now, we are able to describe, in general terms, the process for content discovery on the Web and the consequent URLs recommendations:

1. Let's consider a query $q$ as a list of keywords. We can assume here the query has been tuned by the user for content recall and precision. Recall refers to the set of retrieved documents while precision is the set of relevant documents. Usually these two quantities are inversely proportional: as recall increases, precision decreases.
2. The query is submitted to a set of $t$ global search engines $E_1, \ldots, E_t$ like Google, Yahoo, etc.
3. Each submission will return a search result. Therefore, we will have available $t$ search results $R_1, \ldots, R_t$.
4. Most likely, the search results will have many URLs in common, so we will consider the union $R = \bigcup_{i=1}^{t} R_i = \{u_1 \ldots u_m\}$ representing the $m$ distinct URLs $u_i$.

At this point we are guaranteed to have distinct URLs; however there is a chance some of the URLs might represent the same article. In order to maximize the knowledge around the topic described by the keywords, ideally, we want to select content that is diversified by minimizing document similarity among the retrieved content.

To optimize the similarity matching process we will need to extract the article textual information contained in the HTML page. This is not a trivial operation since the HTML will also contain extraneous content like advertisement, excerpt of other articles, news, etc.

We will present in the "Text Analytics Subsystem" of this paper, a methodology for extracting the article's text from a noisy web page based on an iterative approach that combines paragraphs of text, an operation defined as reduction, until a certain number of reductions have been completed.

5. Let $R = \{u_1 ... u_m\}$ the set of URLs returned by the search engine. We will extract the article information from the URLs $\{u_1 ... u_m\}$ and create the set of documents $D = \{d_1 ... d_m\}$.

6. Let $D = \{d_1 ... d_m\}$ the set of documents extracted by the URLs $\{u_1 ... u_m\}$. We want to define a measure of similarity $S$ so that given two arbitrary documents $d_i$ and $d_j$ we can measure their similarity as $S(d_i, d_j) \in [0,1]$ where 0 is associated with completely dissimilar documents and 1 represents an exact match.

Many measures of document similarity exist today, yet we were unable to find one that would also take into consideration the spatial features of a document. To this end we will define a new similarity measure called Spatial Similarity, based on spatial features, and compare it with the standard Cosine similarity. Moreover, both techniques will be combined into the Spatial Cosine Similarity measure (SCS) that leverages both approaches.

7. Given the documents $\{d_1 ... d_m\}$ we will derive the new set $D' = \{d_1 ... d_v\}$ with v<m (hopefully v<<m) in which $\forall_{i,j} \rightarrow SCS(d_i, d_j) < t$ with $0 \leq i, j \leq v$, $i \neq j$ and $t$ a predefined similarity threshold that we will identify in our experiments and set to an initial value of 0.5.

At this point we are left with a set of URLs $\{u_1 ... u_v\}$, with v<m, representing documents whose content can be assumed to be dissimilar from URLs in the same set.

It would be very useful, before presenting this result to the end user, to provide an ordering of the elements such that document $d_i$ is richer (read "more information") than document $d_j$ for i<j.

One way of achieving this would be to use measures of information like Lexical Diversity or the Information Entropy.

To this end we will introduce the Lexical Entropy $LE(d_i)$ based on term frequencies:

8. Let $LE(d_i)$ be the entropy function associated with document $d_i$.
9. We can induce a total order on $\{d_1, \ldots, d_v\}$ with $LE(d_i)$ so that for i<j $d_i \leq d_j \leftrightarrow LE(d_i) \leq LE(d_j)$ and obtain the ordered set $\{LE_1, \ldots, LE_v\}$ with $LE_1 \geq LE_2 \geq \cdots \geq LE_v$ and $LE_i = LE(d_j)$ for some $j \in [1, v]$, i=1,2,…,u.
10. Finally, the recommendation returned by the system is the set $R_{final} = \{u_1, \ldots, u_v\}$ representing URLs associated to the documents in the ordered set $LE_{final} = \{LE_1, \ldots, LE_v\}$.

## 2. Architecture

We will introduce the overall architecture of our discovery platform. Then, we will look again at each of the processing steps introduced in the previous chapter and we will describe the architectural component that contains their implementation. This will help us to easily identify each module, its function and its flow of information.

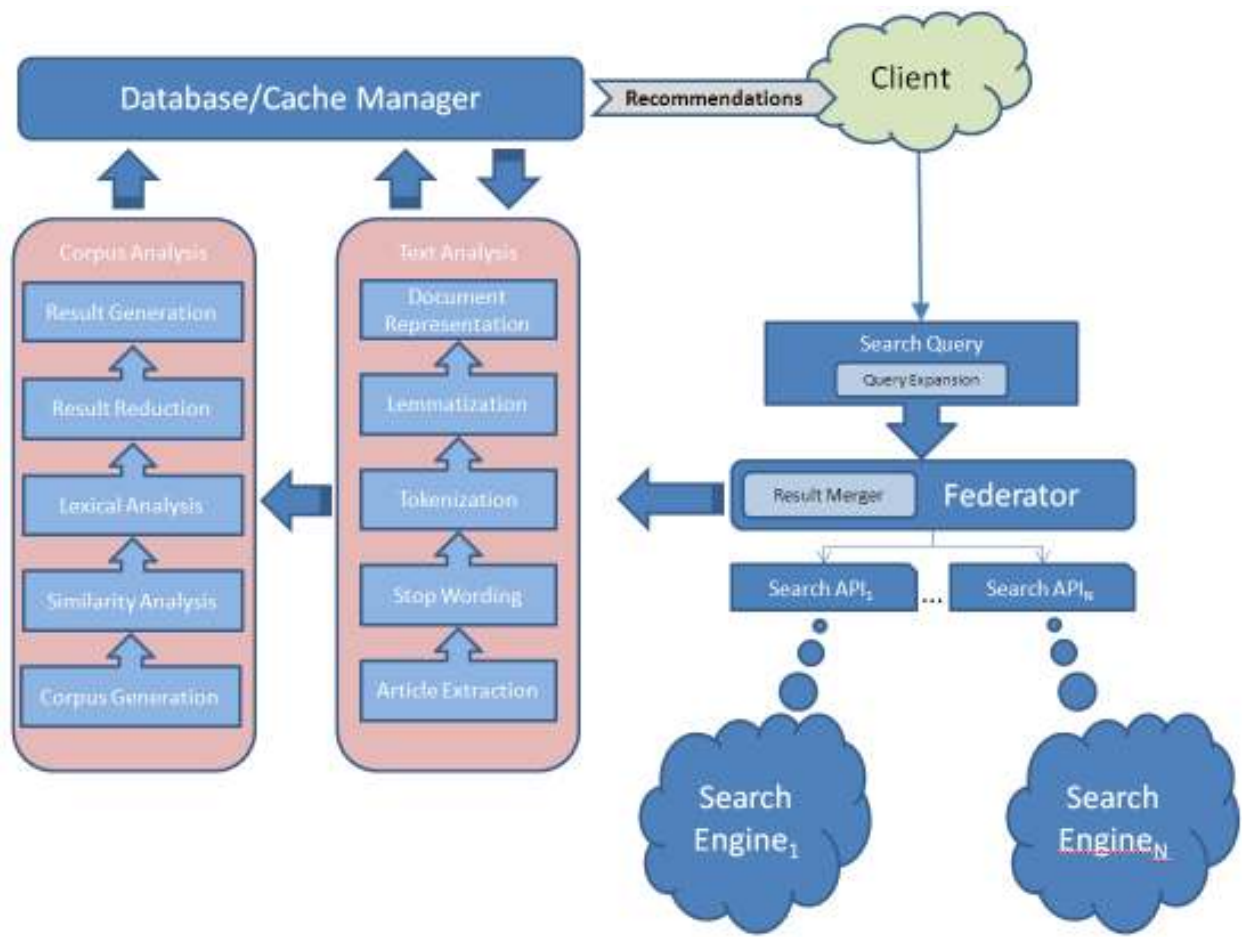

**Fig. 2** – Discovery Platform Architecture.

The overall architecture, shown in Figure 2, is composed of three distinctive modules:

1. **The Search Subsystem**: responsible of taking a query from a client application and federating it to various search engines. The various search results are then merged and made available to the Text Analytic module.

2. **Text Analysis Module**: The search results are processed and the article in each web page is extracted for processing. Stop-wording, tokenization and lemmatization are applied to create a canonical representation of the text.

3. **Corpus Analysis**: Here the set of documents are considered part of a corpus and a similarity and lexical analysis is carried out in order to generate the final recommendation.

During processing the system makes use of a relational database for operations like:

1. Synonyms for query expansion
2. Retrieval of "stop words" vocabulary
3. Tracing of processed URL for duplication detection.
4. Cache management

We will now look into the details and feature of each module.

## 2.1 The Search Subsystem

When a client application submits the query representing the topic of interest the process starts, involving several components as shown in Figure 3.

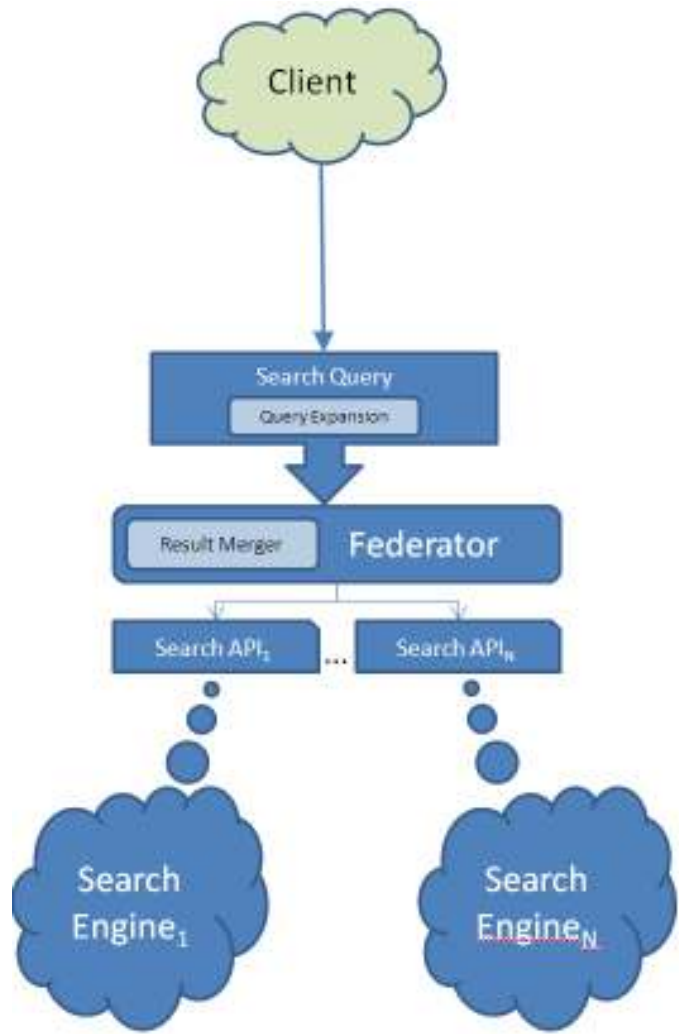

**Fig 3** – Search Components.

A query object is created out of the query specified by the calling application which is then used to perform a federated query using all the available search engines' implementations. In our current implementation we are federating the search query to Microsoft Bing, Bing News, Yahoo and the more specific service Yahoo Blogs, but with the adoption of a factory pattern we can easily create plug-ins for other search engines by simply implementing the *Searchable* interface and adding the class to the appropriate Java package, so that the *SearchEngine* factory will automatically become aware of this new engine implementation.

In Figure 4 we have the class diagram of the Search Subsystem. This is expressed in UML (Unified Language Model) notation which is a standard in object oriented development.

The interface **Searchable** provides two methods signatures:

1. getResults(String)
2. getResult(String,int)

The string in input is the search query and the second method allows specifying the maximum number of results to extract.

In general, for concrete knowledge discovery use cases we would want to enquiry as many search engines as possible in order to maximize document recall, and not sacrifice precision since we don't want to miss any important document when researching critical topics like compliance for example. However, we have to watch out for search queries that return many results (in the order of thousands) because this might be symptomatic of a not well-defined query. By default the system will request 40 documents per query, per engine. The value was chosen based on the fact that any search user will rarely look beyond of the top 20 documents in any search query. With this in mind, if we submit a query to MS Bing and Yahoo you will retrieve a maximum of 80 results.

In order to maximize recall the query is processed for synonym-expansion before being submitted. During this process the query is expanded with a defined list of synonyms for the words it originally contains. For example the term "*automobile*" would not match pages containing the term "*car*" or "*auto*"; so in order to maximize the recall the system will expand this query to "*automobile or car or auto*". The synonyms vocabulary is controlled by the *DictionaryManager* class that provides support for vocabulary-based lists like stop-words and synonyms and is stored in a relational database.

Since each search engine ranks content based on importance, we can assume that such documents are the most important documents on the subject. Obviously, because we do not have control on how the search sites are ranking their indexed documents, it is even more important to have a good selection of engines to query in order to have a good diversity in the way content is ranked.

When the search results from the various search engines are returned, the *SearchResultMerger* class merges the results by adding them to a collection, removing any duplicated content identified with the same URL. This is a simple way of handling duplication when several search engines return the same URL; however, the same content might have been published by different sites and therefore we might still have cases of duplicated content in documents identified by different URLs. We will return to this topic in the section dedicated to Similarity Analysis.

It is important to observe that because the service request is executed with a simple HTTP request without the use of HTTP Headers (protocol used by the browser to send information to the search engine based on system configuration and previous search history), the targeted search service is not able to detect any particular user information and therefore unable to personalize the result based on previous requests. In this way the search result is free from any contextualization/personalization by the search engine that might generate a biased result.

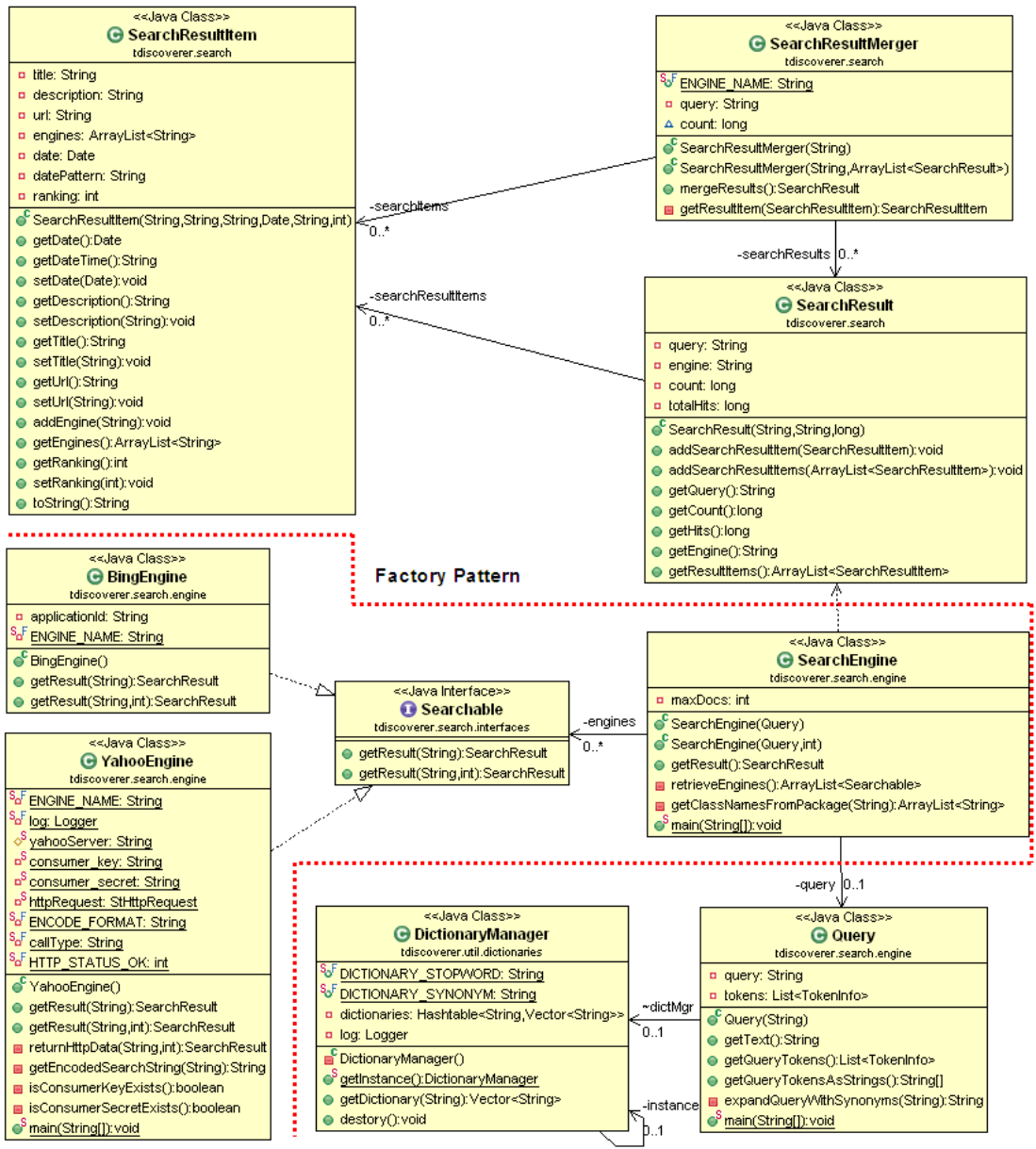

**Fig. 4** – Search Subsystem's UML Diagram.

Merging content resulting from a federated query is an open problem, for which we don't currently have a real solution. The main issue is related to the fact that different engines rank their content in very different ways so it is not possible to compare documents ranked by different search sites in order to have a ranking solution for the complete search result. A very common solution to this problem is to build the merged search result by picking documents in a round-robin fashion from the different results. Other methods do exist of course and might involve semi-supervised learning (SSL), sample-agglomerate fitting estimate or regression methods. Since our research is not focused on the merging of search results, and because we are providing recommendations for topics of interest and not search results in the strict sense, the ranking issue of merged results does not play, in our case, any important role and, therefore, the solution we adopted is to simply shuffle the merged collection of document in a random fashion.

The output provided by the search subsystem will be a collection of *SearchResultItems* objects uniquely identified by their URLs. Moreover, the system also tries to augment each search result with as much metadata as possible with information like title, description, publication date and ranking.

## 2.2 The Web Sampler

The search subsystem can be leveraged so to extract random samples of web pages of any dimension. In fact, in order to test the different modules in the system it would be very helpful to have available a sample of web pages that is randomly generated. The following are good readings on different ways of generating a web sample:

1. C. Snelson, Ed.D., Dec. 2005 – "Sampling the Web: The Development of a Custom Search Tool for Research" – Boise State University.
2. Z. Bar-Yossef, M. Gurevich, Oct. 5$^{th}$, 2008 – "Random Sampling from a Search Engine's Index" – JACM 55(5)

3. M.R. Henzinger, A. Heydon, M. Mitzenmacher, M. Najork – "On Near-Uniform URL Sampling" - http://www9.org/w9cdrom/88/88.html

Despite the methods described in the paper mentioned above, we really wanted to reuse and leverage the search sub-system already developed, and to this end we implemented the following approach:

1. Let *m* be the sample size.
2. From the list of all English nouns, dictionary pick a random set of *m* nouns: $n_1, \ldots, n_m$.
3. For each noun $n_i$ execute a query $q(n_i)$, containing $n_i$ as the sole keyword, against all search engines configured in the system.
4. Pick a single URL from each search result to construct the random sample.

Even though the sample is not uniform across all URLs indexed by the search engines, it contains the randomness necessary for guaranteeing the models do not overfit a particular dataset.

Since the dictionary used by the Web Sampler is comprised of all English nouns, common search results like wekipedia.com, dictionary.com, etc. were removed from the result using a Stop-URL dictionary containing the list of URLs we want to remove from the result.

As an example we ran the Web Sampler to generate a sample of 10 pages and here the results:

```
Generating a sample of size 10
1. Term generation
        poisonous
        cheesecake
        tartan
        tract
        vegetable
        haughty
        trajectory
        outlandish
```

```
    expressive
    fraud

URLs Sampled

    http://kidshealth.chw.edu.au/fact-sheets/poisonous-plants
    http://www.cheesecake.com/History-Of-Cheesecake.asp
    http://www.tartan.nl/index.asp
    http://www.tractbilling.com/?page_id=1529
    http://www.vegetableseeds.net/default.asp
    http://www.haughty.com/about-us
    http://www.addictinggames.com/action-games/trajectory-game.jsp
    http://outlandish.us/Site/The_Company.html
    http://itunes.apple.com/us/app/expressive/id398345416?mt=8
    http://www.fraud.org/welcome.htm
```

## 2.3 Text Analytics Subsystem

Once the set of recommendated candidates have been generated by the Search Subsystem, they will undergo a series of text analytics processes aimed to:

1. Extract the textual article from the web page from each URLs in the search result.

2. Apply stop wording filter to remove frequent terms.

3. Apply lemmatization to standardize the terms to a canonical form.

4. Apply tokenization to allow document analysis in later stages.

The stages are applied in a sequence as show in Figure 5.

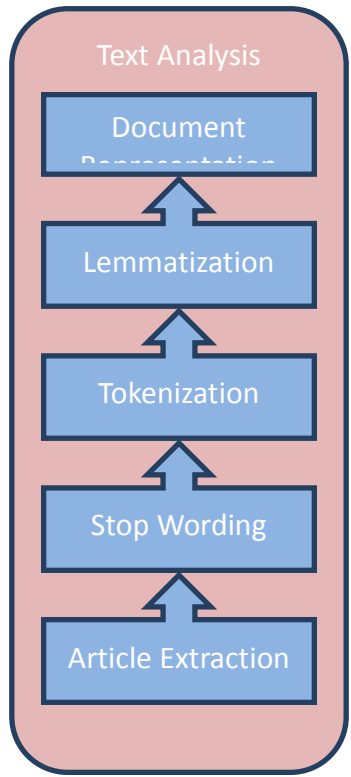

**Fig. 5** – Text Analytic Subsystem.

*Article Text Extractor*

Before performing any meaningful analysis on the corpus of documents associated with the set of URLs returned by the search subsystem, we need to extract the block of text representing the main article from each web document. This is not a trivial task since in a web page the main article can be found along with a variety of extraneous content represented by navigational menus, advertisements, excerpts of news, comments or textual data of other nature as in the following image.

**Fig. 6** – Example of Page with Extraneous Content.

Left untouched the noise represented by the extraneous content has the potential to negatively impact the quality of recommendations provided by the system and must be reduced to acceptable levels.

The problem of extracting textual information from semi-structured HTML content has been well-studied and approaches have been developed to solve this challenge. Existing approaches tend to rely on screen-scraper tools based on manual or tool-assisted generation of rules or code.

These tools tend to perform extremely well when tuned to a particular site, but they lack generalization capabilities when the same set of rules is applied to a new set of pages. Other techniques like Visual Page Segmentation (VIPS) are effective, but computationally expensive.

These techniques have a strategy in common: they rely on the fact that pages within the same web site are based on the same structural template and that we have a finite number of variations between sites. In reality, the order of variation on how a web page is organized is quite large and structure analysis approaches all suffer from the same generalization problem.

Other approaches attempt to induce a wrapper from a limited number of labeled examples, but it is usually limited to a few predefined structures and cannot handle complex or unexpected information structure in the page. Others are more powerful, considering the hierarchical description of the page and fields to be extracted on sample documents in order to induce a set of extraction rules. Unfortunately, while these supervised learning systems spare the effort of an expert writing a wrapper manually, they still require up-to-date, tediously labeled examples for each source and remain difficult to scale. Good insight into this problem can be found in "Extracting Article Text from the Web with Maximum Subsequence Segmentation" by J. Pasternack and D. Roth published in 2009.

Problem statement: we would like to, given an HTML page, extract the contiguous block of text (ideally in its entirety) representing the article and eliminating all the extraneous text.

The process can be described as the following:

1. **HTML Removal**
    a. Consider the HTML content identified by the provided URL.

b. Remove all the HTML code, scripts and comments around textual data and convert markup tags like <tr>, <td>, <p>, <br>, etc. into carriage return and/or new line characters. This was accomplished by using the Apache module Tika. More information can be found in "Apache Tika – A Content Analysis Toolkit" published by the Apache Software Foundation on Nov. 2011.

2. **Article Identification**

    a. Remove all the \r (carriage return) and \t (tabs) characters

    b. Identify all the blocks of text $t_1 ... t_n$ between two consecutive \n (new line) characters as in the following example.

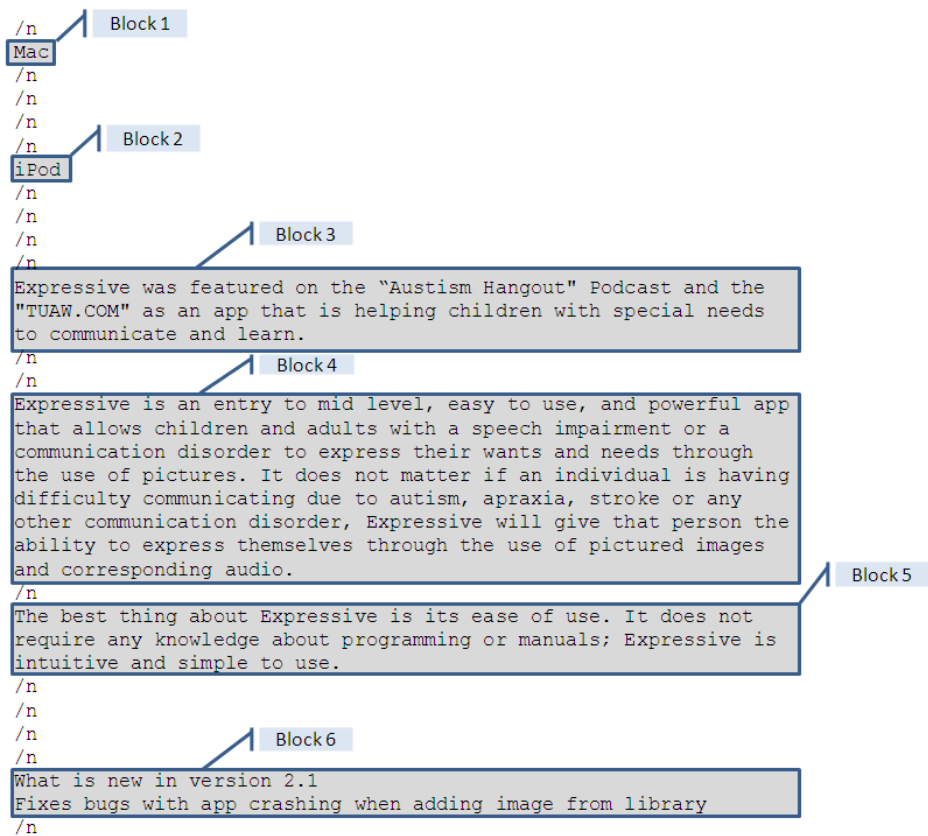

c. Considering the lengths of each block, measured as number of characters, reduce the entire textual content by removing a single /n (new line) character at the beginning of each block. If two blocks were separated by a single /n then merge the blocks.

In the following image you can see the result after one iteration; this is called a single "reduction". Notice Block 4 and Block 5 have been merged in the new Block 4.

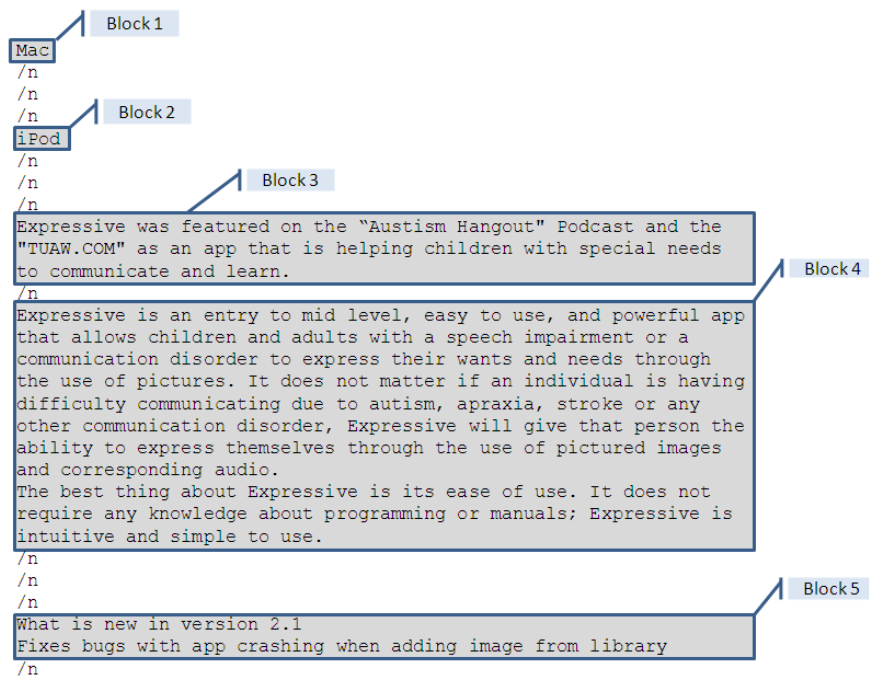

d. Repeat steps (2.b) and (2.c) *k* times.

e. The article is the block of the longest length as shown in the following image.

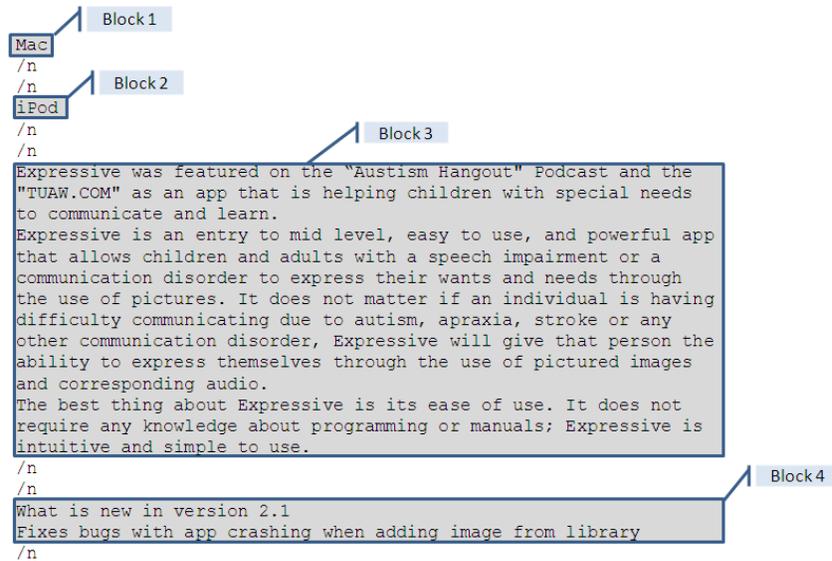

The algorithm is well defined with the exception of the parameter *k*: how many times do we need to iterate through the steps (2.b) and (2.c)?

To answer this question we compared the results of two experiments:

1. Given 10 random samples of web pages, identify *k* as the value that maximized the number of articles extracted using *k* reductions.
2. Set *k* as the mean of the number of reductions executed up to the point when the number of blocks remain the same after two consecutive reductions.

The comparison between the articles extracted by this algorithm with the content on the web pages is a manual comparison.

**Method 1 – Identifying the value of *k***

Since the number of new line characters is limited, the ideal value for *k* is a number between 1 and the maximum number of new line characters between two blocks. After a few experiments we were able to limit our interval to [3, 10] and this was the interval used in our experiment below.

We generated 10 random samples of 10 documents each using the Web Sampler application and for each sample we extracted the articles using a value of *k* varying in the interval [3, 10] and compiled a table as the following:

| Article # | K=3 | K=4 | K=5 | K=6 | K=7 | K=8 | K=9 |
|---|---|---|---|---|---|---|---|
| 1 | 1 | 3 | 3 | 4 | 5 | 5 | 5 |
| 2 | 5 | 5 | 5 | 5 | 5 | 5 | 5 |
| 3 | 2 | 2 | 1 | 2 | 5 | 5 | 5 |
| 4 | 1 | 1 | 2 | 2 | 2 | 2 | 2 |
| 5 | 5 | 5 | 5 | 5 | 5 | 5 | 5 |
| 6 | 5 | 5 | 5 | 5 | 5 | 5 | 5 |
| 7 | 5 | 5 | 5 | 5 | 5 | 5 | 5 |
| 8 | 3 | 3 | 3 | 3 | 3 | 3 | 5 |
| 9 | 5 | 5 | 5 | 5 | 5 | 5 | 5 |
| 10 | 2 | 2 | 2 | 3 | 3 | 4 | 4 |

The values will form a Likert scale that will capture the goodness of fit using the following scale:

**5 - Perfect Match**: the algorithm was able to extract the article content from the web page exactly.

**4 - Strong Partial Match**: the algorithm was able to extract most of the article contained in the web page.

**3 - Partial Match**: the algorithm was able to partially extract the article contained in the web page, but the extracted text either contained extraneous content or was much shorter than the original article.

**2 - Weak Match**: the algorithm was only able to capture a small portion of the article.

**1 - Mismatch**: the algorithm failed to extract the article from the web page.

We collected the results for each of the 10 samples and analyzed the results using IBM SPSS Statistics.

For detailed results of the experiment please refer to Section A.1 "Determining the number of reduction steps" in the appendix.

Looking at the histograms for each of the variables corresponding to value of K from 3 to 9 (K_3, K_4, …, K_9)

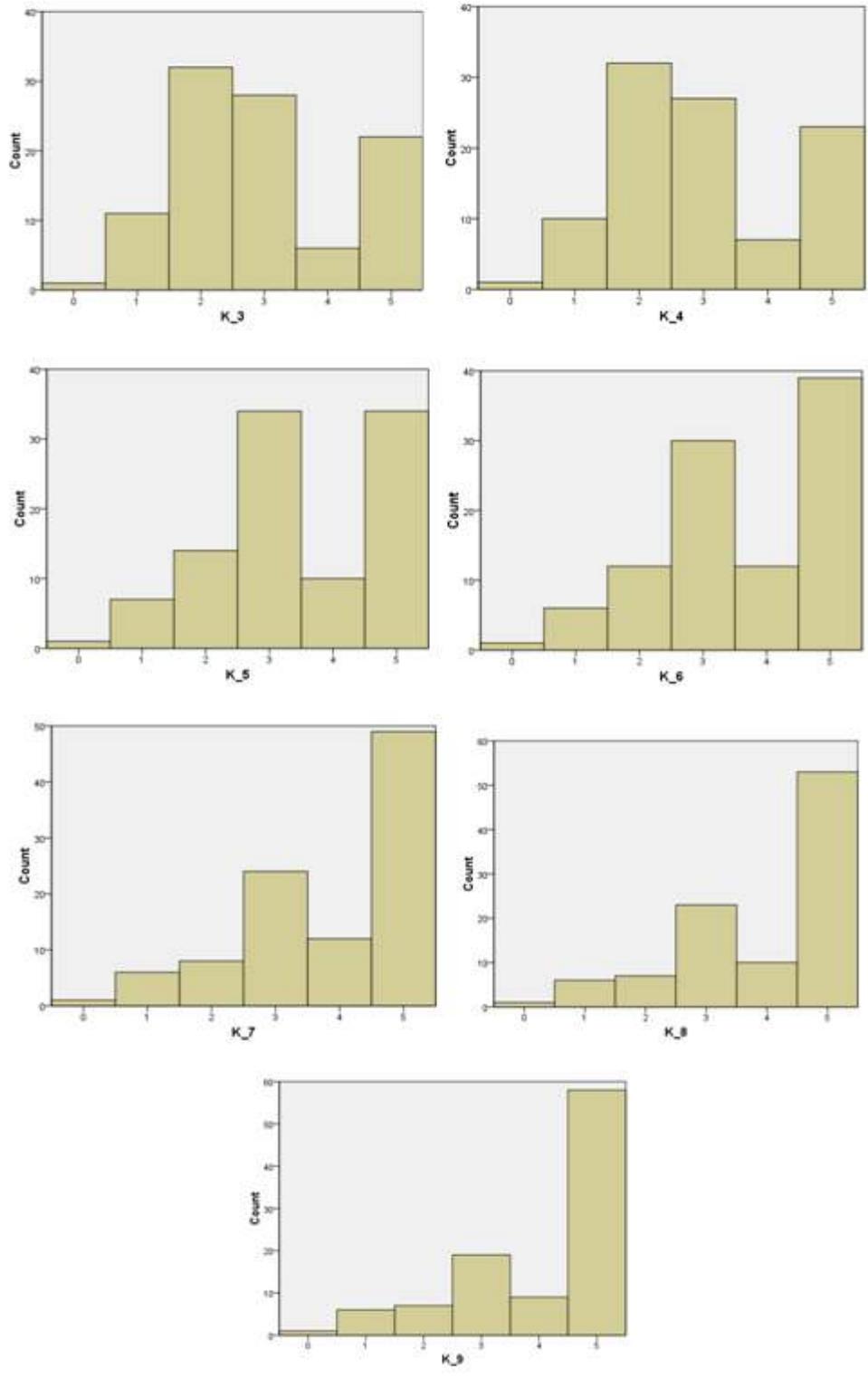

**Fig. 7** – Histograms of the Extraction Quality for Different Values of K.

As we can see as the value of K increases so does the quality of the extraction until it plateaus at *K=9*.

In terms of accuracy, as reported in Table 1, we have an accuracy of 67 % when considering high quality results like *Perfect Match* or *Strong Partial Match*. Since the extraction of a good portion of the article will suffice for our purposes, we can loosen our quality requirements a bit and also include *Partial Match* in our category of acceptable results. This increases the accuracy level to a healthy 86%.

| K_9 | | | | |
|---|---|---|---|---|
| | Frequency | Percent | Valid Percent | Cumulative |
| 1 | 7 | 7.0 | 7.0 | 7.0 |
| 2 | 7 | 7.0 | 7.0 | 14.0 |
| 3 | 19 | 19.0 | 19.0 | 33.0 |
| 4 | 9 | 9.0 | 9.0 | 42.0 |
| 5 | 58 | 58.0 | 58.0 | 100.0 |
| Total | 100 | 100.0 | 100.0 | |

**Table 1** – Frequencies *K=9*.

This method identifies 9 as the number of reductions necessary for extracting an article with 86% accuracy rate and with a quality lower-bound of "Partial Match."

**Method 2 – Select *K* as the average number of reductions.**

In this case we executed the entity extraction algorithms on 100 random web pages generated by the *Web Sampler* application. We recorded the number of reductions necessary until the number of blocks remained the same within two consecutive reductions and calculated the average as reported in Table 2.

Since the 100 executions have reduced the textual information an average of 26 times we could conclude that K=26 is the value we are looking for. However, we also notice that the median is 15, not too far from K=9 selected by the previous method. If we also look at the frequency values we can see that lower values of K are used more frequently as shown in Figure 8 (a). If we zoom-in on the frequency graph as in Figure 8 (b) and also look at Table 2, we can see three peaks that are associated with values of $K$ equal to 6, 8 and 14.

We can exclude $K=6$ as a solution from the previous test so we can restrict $K$ to a value between 8 and 14.

For values of K>9, the quality of the text extracted from the pages starts deteriorating due to the inclusion of extraneous content so we therefore set the value of $K$ to 9.

**Reductions**

|       |       | Frequency | Percent | Valid Percent | Cumulative Percent |
|-------|-------|-----------|---------|---------------|--------------------|
| Valid | 5.00  | 5         | 5.0     | 5.0           | 5.0                |
|       | 6.00  | 7         | 7.0     | 7.0           | 12.0               |
|       | 7.00  | 3         | 3.0     | 3.0           | 15.0               |
|       | 8.00  | 8         | 8.0     | 8.0           | 23.0               |
|       | 9.00  | 5         | 5.0     | 5.0           | 28.0               |
|       | 10.00 | 4         | 4.0     | 4.0           | 32.0               |
|       | 11.00 | 1         | 1.0     | 1.0           | 33.0               |
|       | 12.00 | 5         | 5.0     | 5.0           | 38.0               |
|       | 13.00 | 3         | 3.0     | 3.0           | 41.0               |
|       | 14.00 | 7         | 7.0     | 7.0           | 48.0               |
|       | 15.00 | 2         | 2.0     | 2.0           | 50.0               |
|       | 16.00 | 5         | 5.0     | 5.0           | 55.0               |

**Descriptive Statistics**

|            | N   | Min. | Max.   | Mean    | Median  | Std. Deviation |
|------------|-----|------|--------|---------|---------|----------------|
| Reductions | 100 | 5.00 | 120.00 | 25.6600 | 15.5000 | 23.27231       |

**Table 2** – Summary Statistic for the Number of Reductions Executed by the Algorithm.

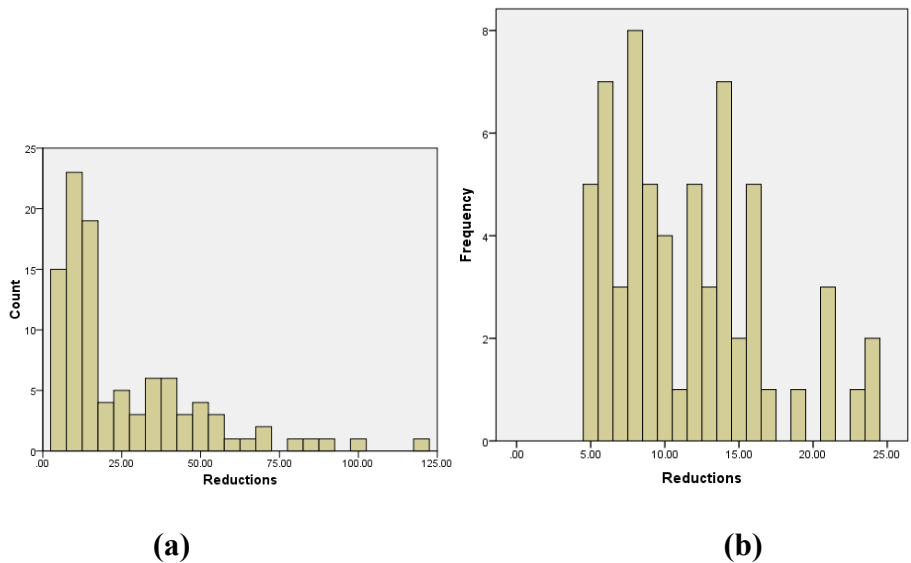

(a)            (b)

**Fig. 8** – Histogram for the Number of Reductions Executed by the Algorithm.

*Stop Wording*

Many factors affect the result of the textual analysis of a document. In general, it is impacted by its content, the language in which the content is written, the degree to which lemmatization is employed, the extent to which stop words are removed, and the degree to which synonym expansion using thesauri are utilized.

*Stop-Words* are words which are frequent, but have little meaning within the context of the document, for example the articles "the", "a", "an" and can, therefore, be eliminated. We used the same stop-word dictionary containing 127 terms used by the NLTK python package which seems to offer a good balance of term parsimony.

This operation is executed by the *Lemmatizer* class by invoking the *removeStopWords* method asspecified in the class diagram in Figure 9.

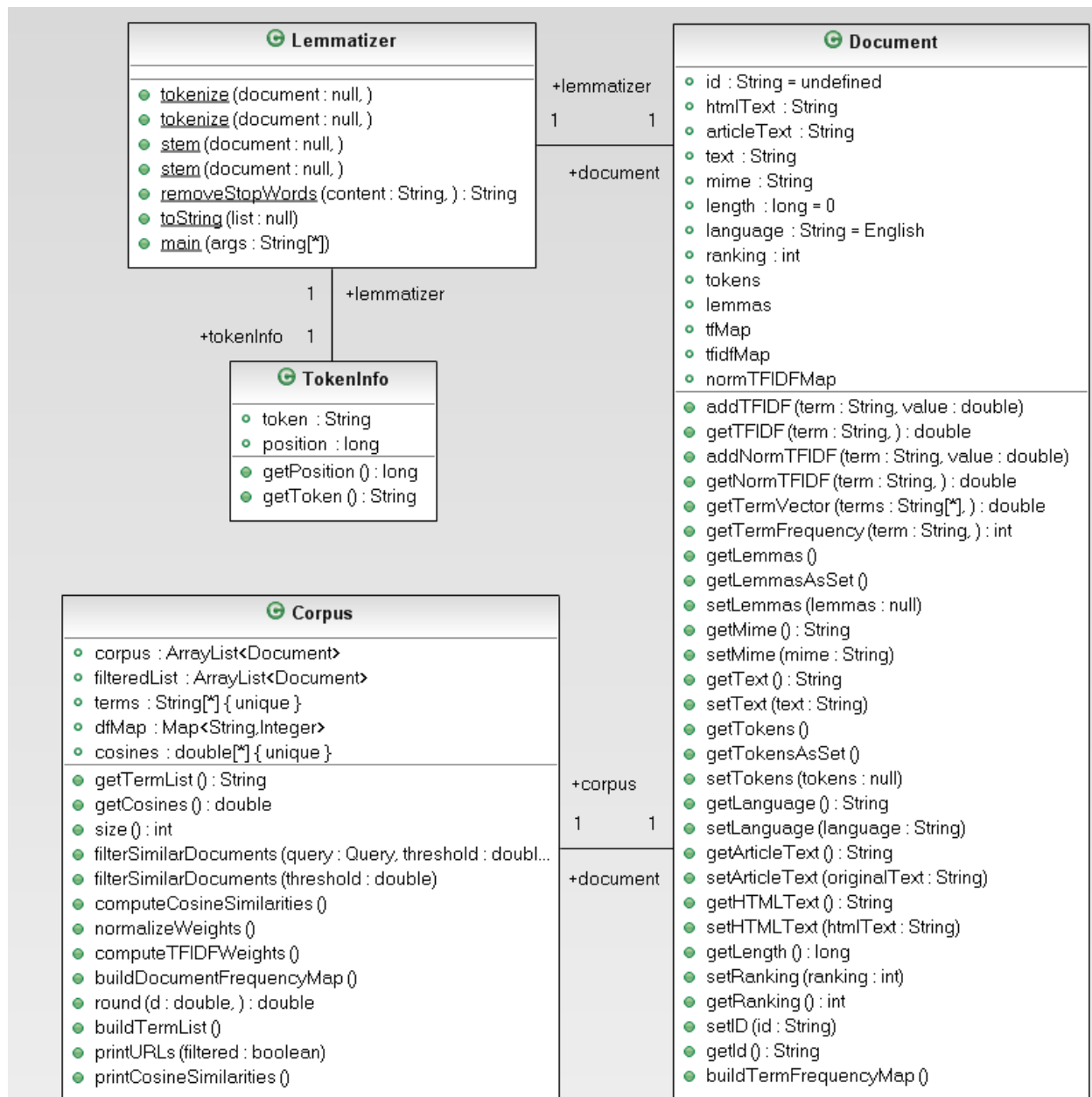

**Fig 9** – Text Analysis Subsystem UML Diagram.

*Tokenization*

The next step is tokenization which is part of linguistic analysis, where text is split into word entities. This procedure is accomplished with the use of the Stanford Language Processing library that consist of a set of a natural language analysis tools which can manipulate an English document as input and provide the base forms of words, their parts of speech and basic annotation capabilities like company names, people, date, etc. Moreover it can provide the markup structure of sentences in terms of phrases and word dependencies and indicate which noun phrases refer to the same entities. You can find more information on this library at [http://nlp.stanford.edu/software/corenlp.shtml](http://nlp.stanford.edu/software/corenlp.shtml), the home of the Stanford Natural Language Processing Group.

The resulting tokens are stored in a list of *TokenInfo* objects in which we also store location information on where the token is positioned within the text. During the tokenization phase a token is also processed against the synonym dictionary that will perform a lookup and return the base synonym in case a set of synonyms is found. The base synonym, or preferred term, is defined as the base word returned by the system when a token matches a synonym group and it is always the same, independent of which word matches the group.

The synonyms process has been implemented using the Wordnik API for synonym lookup as described at [www.wordnik.com](www.wordnik.com) in the Wordnik Services area.

### *Stemming Vs. Lemmatization*

Because a word might appear in the text in many forms (singular or plural for example) it is important to transform all tokens in their base-form, a procedure called *Stemming*. This algorithm-based procedure is quite fast and applies linguistic rules for each token in order to transform it into the base form. The resulting text will contain words in their canonical (base) form and it is now ready for the application of analytical processes like words counts, similarity, etc.

Lemmatization is a more complex operation.

As with the stemming technique, the goal of lemmatization is to represent words in their canonical forms, but unlike stemming that considers single words, lemmatization takes into consideration the context in which the words appear. Consider the word "meeting." It can be a verb or a noun, a distinction that is completely missed by the stemming, but taken into consideration during lemmatization. In this case a stemmer will produce "meet" in both case, while a lemmatizer will produce "meet" in the case of a verb and "meeting" in the case of a noun. Another example is the word "worst" that a lemmatizer will transform into "bad", a link that will be completely missed by a stemmer.

Because lemmatization will require a combination of text processing and database lookup, its computational overhead is higher than stemming, moreover it usually requires the text to be processed with a POS (Part Of Speech) tagger that provides the contexts around each token that will be later processed by the lemmatizer.

In deciding whether to apply lemmatization instead of the simpler and faster stemming, we researched the topic and did not find a real strong argument for lemmatization in applications outside the realm of search engines. In fact, the increase in precision using lemmatization over stemming is very modest, on the order of few percent points. As for the ability to reduce the number of distinct words, the two methods are once again very comparable as shown in Table 3. A good reading on this topic can be found in "An Evaluation of Linguistacally-motivated Indexing Schemes" by A. Arampatzis, T. van der Weide, C.H.A. Koster and P. van Bommel in 2000, published by the Proceedings of BSC-IRSG 2000 Colloquium on IR Research.

| Method | Distinct Terms | Reduction |
|---|---|---|
| Tokenization Only | 34,030 | Baseline |
| Stemming | 27,205 | 20.0% |
| Lemmatization | 26,952 | 20.8% |

**Table 3** – Distinct Term Occurrences.

The stemming procedure is carried out by the *Lemmatizer* class through the method *stem* as shown in Figure 9. The method can take either a document or a list of tokens and return a list of *TokenInfo* objects containing the base form of the respective token.

*Document Representation*

At the end of the text processing pipeline, the resulting content and metadata is stored in an instance of the *Document* class. This class, besides storing the original and the transformed version of the text, contains a rich set of metadata information:

- Language
- Text length
- Mime type
- Document ranking (from the search result)
- Text frequencies information
- Tokens position within the text and their length

Regarding to frequencies, the *Document* object will contain the term frequency $tf_{t,d}$ which is defined as the number of occurrences of the term $t$ in the document $d$.

Once the full set of documents associated with the search result coming from the search subsystem has been retrieved and processed, the system is ready to proceed to the analysis of the corpus of documents.

**Corpus Analysis Subsystem**

As we have seen, the articles retrieved by the search subsystem are processed and collected in a corpus of documents representing the collection of the most current articles around the topic specified by search keywords.

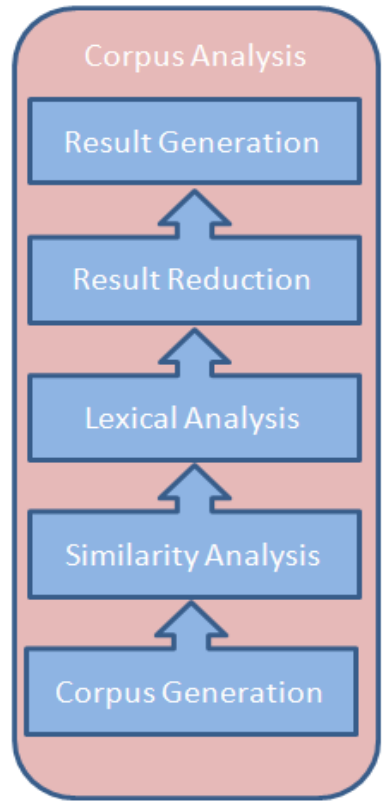

**Fig. 10** – Corpus Analysis Subsystem.

With the corpus in place we can now carry out a series of analyses as described in the flow in Figure 10.

The main goal of this process is to reduce the number of documents by eliminating:

1. **Duplication**: different web sites might publish the same article.
2. **Similarity**: different articles might be very similar and therefore provide the same amount of information.

Once the reduction is complete the system will provide the result to the user as the final recommendation on the topic.

## *Corpus Generation*

The Text Analysis subsystem provided, at the end of its process, an array of *Document* objects. This array is fed into the constructor of the *Corpus* class to generate our corpus.

In order to identify relevant terms, several frequency measures are calculated during the construction of the corpus:

1. Construction of the corpus terms list using the full set of tokens extracted in each document.
2. Calculation of the document frequencies $df_t$ defined as the total number of documents in the corpus containing the term *t*. The reasoning behind this measure is the fact that it is better to use document-level statistics (such as the number of documents containing the term) than it is to use collection-wide statistics for the term.
3. Calculation of the Inverse Document Frequency $idf_t = \log \frac{N}{df_t}$.
4. Computation of the *tf-idf* weights defined as *tf-idf* $= tf_t \cdot idf_t$ as described in "Practical Text Mining with Perl" written by R. Bilisoly in 2008 and published by Wiley Publications.

The $tf_t - idf_t$ assigns to the term *t* a weight in the document *d* which is:

1. Highest when the term *t* occurs many times within a small number of documents.
2. Lower when the term occurs fewer times in a document, or occurs in many documents.
3. Lowest when the term occurs in virtually all documents.

These weights will play an important role in the determination of a measure of similarity between the documents that we will address in the next section.

### *Similarity Analysis*

#### Cosine Similarity

Each document *d* in the corpus can be represented as a vector in the hyperspace with a vector component for each dictionary term. The components are computed using the ***tf-idf*** weights scheme we described in the previous section.

Using this representation, called the *Vector Space Model*, each document in the corpus can be viewed as a vector in the hyperspace in which each term is associated to a dimension. The dimension of the hyperspace is equal to the total number of terms in the corpus and when a document does not contain a particular term the value for the associated component is zero.

With this model in place, if two documents are "similar" they will contain related terms and we can expect their vectors to be "close" in the hyperspace. By close we mean the angle between the two vectors is very small.

Since the relative distribution of terms might be identical in two documents, but the absolute term frequencies of one may be much larger in one than the other, the correspondent vectors might have different lengths. To solve this problem we will normalize the vectors so that each document will be represented by unit vectors.

Let $d_1$ and $d_2$ be represented by the two standardized vectors $\vec{v}_1$ and $\vec{v}_2$ where the similarity between the two is the cosine of the angle between them as shown in Figure 11.

More formally we have

$$sim(d_1, d_2) = \vec{v}(d_1) \cdot \vec{v}(d_2) = \cos\theta$$

When the two documents differ in each dimension they are represented by two orthogonal vectors and therefore the angle between them is $\theta = 90\ degree$, consequently $\cos 90° = 0$.

On the other hand, if two documents are exactly the same then the angle between their vector representations is 0 and $\cos 0° = 1$; therefore the similarity between two documents is a value between 0 and 1.

Because this measure of similarity is defined over the cosine function it is called **Cosine Similarity** and it is currently one of the most utilized measures of similarity in the industry.

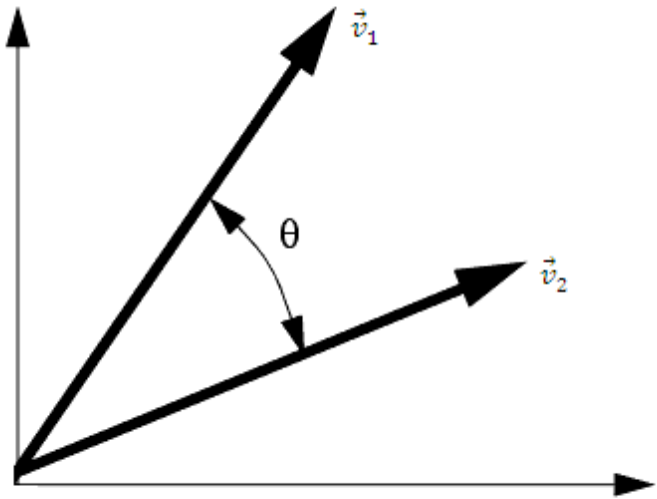

**Figure 11** – Angle between two document vectors

Spatial Similarity

The cosine similarity, defined in the previous section, represents a widely used standard in the information industry, one that's implemented in many commercial search engines like Microsoft Fast, Endeca, Autonomy and in the open source Apache Solr, to only mention the most recognizable names.

Despite its popularity, the cosine similarity has the drawback of not considering word placement in the text under analysis. A classic example of this problem is the comparison of the two texts "John loves Mary" and "Mary loves John". These two simple sentences use the same words, but their meaning is completely different. When comparing these documents, the cosine similarity will produce the exact-match value of 1, due to identical term vectors, yet this result is arguable to say the least.

To overcome this lack of spatial analysis we introduce a new measure of similarity: Spatial Similarity.

Let $d_i$ and $d_j$ be two documents and let $p_i^{t^k}$ and $p_j^{t^k}$ be the ordinal position of the k-instance of the term $t$ (starting from 0) in the document $d_i$ and $d_j$ respectively when the k-instance of the term $t$ appears in both documents, 0 otherwise.

We define the *Spatial Difference* for the term $t$ as the quantity $sd_{i,j}^t = \sum_k \frac{|p_i^{t^k} - p_j^{t^k}|}{p_i^{t^k} + p_j^{t^k}}$.

Each term in the summation can be 0 when the *k*-instance of the term *t* appears in exactly the same ordinal position in both documents and 1 when *t* is in the initial position of 0 in $d_i$ and in some other position $p_j^t \neq 0$ in $d_j$ (or vice versa). Therefore the quantity $\frac{|p_i^{t^k} - p_j^{t^k}|}{p_i^{t^k} + p_j^{t^k}} \in [0,1]$.

Finally we define the Spatial Similarity of the two documents $d_i$ and $d_j$ to be the quantity:

$$ss(d_i, d_j) = \frac{\sum_t sd_{i,j}^t}{\lambda}$$

In the numerator we have the sum of the Spatial Differences $sd_{i,j}^t$ between documents $d_i$ and $d_j$ for all common terms *t*, and at denominator we have the number of matches $\lambda$.

Because the numerator is the summation of quantities between [0,1] appearing no more than $\lambda$ times, the largest value $SS(d_i, d_j)$ can assume is $\frac{1+1+\cdots+1}{\lambda \text{ times}} = \frac{\lambda}{\lambda} = 1$ and a smallest value of 0 when two documents are exactly the same. Therefore $0 \leq SS(d_i, d_j) \leq 1 \ \forall \ d_i, d_j$.

In order for $SS(d_i, d_j)$ to have the same direction as other measure of similarity (1 for an exact match and 0 for total dissimilarity) we modify its definition in its final form:

$$SS(d_i, d_j) = 1 - \frac{\sum_t sd_{i,j}^t}{\lambda}$$

If we consider the "John loves Mary" and "Mary loves John" example we have:

$$SS(\text{"John loves Mary"}, \text{"Mary loves John"}) = 1 - \frac{\sum_t sd_{i,j}^t}{\lambda} = 1 - \frac{\frac{|0-2|}{2} + \frac{|1-1|}{2} + \frac{|2-0|}{2}}{3}$$

$$= 1 - \frac{2}{3} = 0.33$$

This result is quite different from the exact match produced by the cosine similarity and better captures the spatial differences between the two texts.

Spatial Cosine Similarity

In order to consider both spatial and word features of documents we can combine the two similarities using a weighted approach and finally we can define the Spatial Cosine Similarity between documents $d_i$ and $d_j$ as:

$$SCS(d_i, d_j) = \alpha * sim(d_1, d_2) + (1 - \alpha) * SS(d_i, d_j)$$

with $\alpha \in [0,1]$. A value of 0 will make SCS coincide with the spatial similarity $SCS(d_i, d_j) = SS(d_i, d_j)$ and a value of 1 will make SCS coincide with the cosine similarity $SCS(d_i, d_j) = sim(d_i, d_j)$.

Because we have articles that might contain extraneous content after the automatic extraction process, the spatial information might not be accurate and therefore we will consider the cosine similarity slightly more important by picking a value of $\alpha = 0.6$.

Considering, once again, the previous example we have:

$$SCS(\text{"John loves Mary"}, \text{"Mary loves John"}) = 0.6 * 1 + 0.4 * 0.33 = 0.73.$$

This value indicates a good balance between the undeniable similarity of the texts and their spatial differences.

Following this strategy we have two questions that we need to answer:

1. What is the threshold to use to determine when two documents are similar?
2. If two documents are similar which one should we eliminate?

### Similarity Threshold

Since the measure of similarity $SCS(d_i, d_j)$ still depends on the size of the corpus and on the frequency of terms across documents, it is not a trivial matter to identify a similarity threshold that will tell us when two documents are similar. In fact, the document frequencies will tend to decrease as the size and number of documents in the corpus increases, since the probabilities of the same term appearing on different documents will increase as well.

In order to have a sense of how the spatial cosine similarity value for two sets of seeded documents changes within a corpora we setup a small experiment in which the similarity of the two sets of seeded documents are measured as we add random documents to the corpora. See Appendix B for more information.

The first set of seeds is comprised of two documents represented by the same article in which we changed very few terms. This has a very high similarity value. The second set of seeds is represented by two documents taken from two different newspapers talking about a related subject. Such documents would be considered similar to a human reader. The results are shown in Tables 4 (a) and 4 (b) in which we also compared the influence of the corpus size for both measures of similarity: Cosine and SCS.

The experiment was repeated several times with analogous results.

The SCS also showed a lesser influence to the corpus size with smaller variations in the results ranging from 0.03/0.7. Comparing this with the Cosine function, that has variations as large as 0.12, we can safely leave the similarity threshold at the default value of $\sigma = 0.5$.

| Size of Corpus | Similarity of Set #1 | Similarity of Set #2 |
|---|---|---|
| 4 | 0.89 | 0.52 |
| 5 | 0.89 | 0.53 |
| 10 | 0.90 | 0.54 |
| 15 | 0.90 | 0.54 |
| 20 | 0.91 | 0.56 |
| 30 | 0.92 | 0.57 |
| 40 | 0.92 | 0.59 |

**Table 4 (a)** – Similarity Variations with Different Corpus Sizes Using Spatial Cosine Similarity.

| Size of Corpus | Similarity of Set #1 | Similarity of Set #2 |
|---|---|---|
| 4 | 0.85 | 0.48 |
| 5 | 0.85 | 0.50 |
| 10 | 0.87 | 0.51 |
| 15 | 0.86 | 0.52 |
| 20 | 0.89 | 0.44 |
| 30 | 0.90 | 0.57 |
| 40 | 0.91 | 0.60 |

**Table 4 (b)** – Similarity Variations with Different Corpus Sizes Using Cosine Similarity.

When two documents are similar, that is $SCS(d_i, d_j) > 0.5$, the amount of information provided by one of the two documents is redundant and can be eliminated from the corpus since it does not add any knowledge or adds a negligible amount. Therefore, during the Similarity Analysis we will remove all documents that have a similarity greater than $\sigma = 0.5$.

Obviously, in case of a similarity match we cannot remove both documents, else the knowledge associated to their information will be lost, and we are faced with the question of: "Which one of the two documents, determined to be similar, should we remove?"

To help answer this question we use another piece of information represented by the query terms used in retrieving the content.

Let $d_i$ and $d_j$ be two documents such as $SCS(d_i, d_j) > \sigma$ and let the terms $q_1 \ldots q_n$ be the terms used in the query.

We can calculate the average number of times the terms in the query appear in document as:

$$\mu_i = \frac{\sum_{k=1}^{n} f_{d_i, q_k}}{\# \text{ terms in } d_i} \text{ for } i = 1,2$$

with $f_{d_i, q_k}$ being the number of times the query term $q_k$ appears in document $d_i$.

Since we would like to have terms that are evenly matched in the document, we can now calculate the dispersion of the query terms around their averages and retain the document that minimizes this quantity.

Let $s_i = \frac{\sum_{k=1}^{n}(f_{d_i,q_k} - \mu_i)^2}{\# \ terms \ in \ d_i}$ $for \ i = 1,2$ be the dispersion of the query terms around the average for the document $d_j$.

We will retain the document $d_j$ if $s_j = \min(s_i, s_j)$, otherwise we will retain $d_i$.

### *Lexical Entropy*

One of the most intuitive measurements that can be applied to unstructured text is lexical diversity. In formal terms this measure describes the range of a speaker's vocabulary and could be used as a way of ranking the remaining documents in the corpus.

Lexical diversity is related with the range of vocabulary used in the article, so that a document containing $T + \Delta$ different words it is said to be more diverse than one with just $T$ as it was nicely described in the "Developmental Trends in Lexical Diversity" study released in 2004 by P. Duran, D. Malvern and B. Richards from the Oxford University Press.

In order to normalize the lexical diversity measure to be document independent we will use a ratio defined as $L(d) = {T_d}/{N_d}$ in which $T_d$ represents the number of unique tokens in $d$ and $N_d$ is the total number of tokens.

If we have a document in which all tokens are different we have $L(d) = 1$, whereas if the document is comprised of a repeated token then $L(d) = \frac{1}{N_d}$. Therefore, Lexical Diversity $L(d)$ is a number between $(0,1]$.

The ratio $T_d/N_d$ has been the subject of many speculations especially to the fact that this measure might be very sensitive to the age range and education level of the writer. However, since we are considering articles published over the internet we will assume authors in their adulthood and with middle-to-high level of education. A good reference to this issue can be found in "Lexical Diversity and lexical density in speech and writing: a developmental perspective" by V. Johansson and published in 2008 by Lund University.

Another useful piece of information is the document Entropy which measures the richness of a document and is defined as

$$E_d = \frac{1}{\lambda}\sum_{i=1}^{n} tf_i \left(\log_{10}(\lambda) - \log_{10}(tf_i)\right).$$

with $\lambda$ representing the total number of terms in the document taken from a set of $n$ distinct terms and $tf_i$ being the term frequency of the $i$-th term in the same document. We used the definition described in the web article "Entropy Text Analyzer" at http://miup2002.fc.ul.pt/problemas/I/I.html from the MIUP 2002 challenge.

Since the maximum entropy for a test is obtain from a text in which each work occurs exactly once, $\lambda = n,$ then we have

$$E_{max} = \frac{1}{\lambda}\sum_{i=1}^{n} 1 \left(\log_{10}(\lambda) - \log_{10}(1)\right) = \frac{\lambda * \log_{10}(\lambda)}{\lambda} = \log_{10}(\lambda)$$

To compare texts with a different number of words $\lambda$, we define the relative entropy measure:

$$E_{rel}^d = \frac{E_d}{E_{max}}$$

We can now define the Lexical Entropy $LE_d$ as the weighted quantity:

$$LE_d = \alpha * L(d) + (1 - \alpha) E_{rel}^d$$

Because we are more interested in the richness of a document, we will choose a value of $\alpha = 0.4$ resulting in a Lexical Entropy of

$$LE_d = 0.4 * L(d) + 0.6\, E_{rel}^d$$

in which the document entropy plays a more important role than the lexical diversity.

### *Result Generation*

We recall that during the similarity analysis the original corpus $D = \{d_1 \ldots d_m\}$ was, eventually, reduced by eliminating all similar documents and contained the remaining $D' = \{d_1 \ldots d_v\}$ documents with $v \leq m$.

The lexical entropy can impose a total ordering on $\{d_1 \ldots d_v\}$ by calculating the function $\boldsymbol{LE_d}$ for each document in $D'$. This will generate a new set $D'' = \{d'_1, \ldots, d'_v\}$, different from the original, in which the elements have been ordered in the descending order of their lexical entropy, that is $LE_{d_1} \geq LE_{d_2} \geq \cdots \geq LE_{d_v}$.

The final recommendations provided by the system will be the set of URLs associated with the documents in $D''$. The ordering of the elements is important since the receiving system will be able to discriminate between the URLs if it will only need to utilize a subset of them.

## 3. Real Examples for Topical Discovery

### 3.1 Discovery of topical content on "Innovation in Korea"

Being responsible for company innovation in Asia, we would like to discover interesting content about innovation in Korea. In order to provide a nice format for the output, we will reduce the number of documents returned from the various search engines to 4.

This can be achieved with 4 lines of Java code as in the following example:

```
1. Query query = new Query("innovation and korea");
2. WebDiscoverer discoverer = new WebDiscoverer();
3. discoverer.setMaxDocuments(4);
4. discoverer.discoverContent(query);
```

In this case we've created a new query in which we've specified the search query "innovation and korea" representing the search context. Next we've created an instance of the WebDiscoverer and set the maximum number of documents to be returned from the various search engines to 4.

Lastly we called the discovererContent method with the query object and the system activates the discovery process described in this paper, producing the following output.

```
Querying: Bing
Retrieved 4 documents.

Querying: Blogs
Retrieved 4 documents.

Querying: Yahoo
Retrieved 4 documents.

Querying phase for "innovation and korea" completed.

Error retrieving one of the specified URL
Search subsystem retrieved a total of 11 documents.
Calculating Document Similarities on 11 documents.

SPATIAL COSINE SIMILARITY TABLE ----------
```

```
1 - http://ukinrok.fco.gov.uk/en/about-us/working-with-korea/science-
innovation/science-and-innovation-in-korea
2 - http://www.huffingtonpost.com/c-m-rubin/the-global-search-for-
edu_34_b_1407464.html
3 - http://edunloaded.blogspot.com/2009/06/korea-technological-innovations-
2.html
4 - http://pressnews.com/articles/2012/04/07/dayton/05reppaulsentalkwpic.txt
5 - http://edunloaded.blogspot.com/2009/05/new-innovations-in-korea.html
6 - http://candicelanier.blogspot.com/2012/01/s-koreas-different-path-to-
technology.html
7 - http://www.issues.org/24.1/chung.html
8 - http://www.prweb.com/releases/2012/4/prweb9369229.htm
9 - http://innovationkorea.eventbrite.com/
10 - http://bennosgg.wordpress.com/2010/02/17/export-oriented-economic-
models-and-their-effect-on-education
11 - http://news.yahoo.com/pwn-technologies-wins-iwa-project-innovation-
award-2012-080040615.html

 1 - *1.00   0.27   0.21   0.26   0.18   0.30   0.42   0.30   0.23   0.30   0.29
 2 -  0.00  *1.00   0.15   0.28   0.16   0.21   0.25   0.22   0.16   0.25   0.22
 3 -  0.00   0.00  *1.00   0.18   0.25   0.19   0.14   0.13   0.23   0.19   0.12
 4 -  0.00   0.00   0.00  *1.00   0.19   0.24   0.25   0.25   0.21   0.26   0.25
 5 -  0.00   0.00   0.00   0.00  *1.00   0.15   0.19   0.13   0.31   0.24   0.12
 6 -  0.00   0.00   0.00   0.00   0.00  *1.00   0.28   0.25   0.26   0.28   0.25
 7 -  0.00   0.00   0.00   0.00   0.00   0.00  *1.00   0.26   0.16   0.31   0.26
 8 -  0.00   0.00   0.00   0.00   0.00   0.00   0.00  *1.00   0.13   0.25  *0.98
 9 -  0.00   0.00   0.00   0.00   0.00   0.00   0.00   0.00  *1.00   0.19   0.12
10 -  0.00   0.00   0.00   0.00   0.00   0.00   0.00   0.00   0.00  *1.00   0.26
11 -  0.00   0.00   0.00   0.00   0.00   0.00   0.00   0.00   0.00   0.00  *1.00

Calculating Document Similarities on 11 documents
Calculating Document Similarities on 10 documents
The corpus size after filtering is of 10 documents.

- Selected Documents -----------------------------------
1 - http://www.issues.org/24.1/chung.html
2 - http://www.huffingtonpost.com/c-m-rubin/the-global-search-for-
edu_34_b_1407464.html
3 - http://ukinrok.fco.gov.uk/en/about-us/working-with-korea/science-
innovation/science-and-innovation-in-korea
4 - http://news.yahoo.com/pwn-technologies-wins-iwa-project-innovation-award-
2012-080040615.html
5 - http://pressnews.com/articles/2012/04/07/dayton/05reppaulsentalkwpic.txt
6 - http://candicelanier.blogspot.com/2012/01/s-koreas-different-path-to-
technology.html
7 - http://bennosgg.wordpress.com/2010/02/17/export-oriented-economic-models-
and-their-effect-on-education/
8 - http://edunloaded.blogspot.com/2009/05/new-innovations-in-korea.html
9 - http://edunloaded.blogspot.com/2009/06/korea-technological-innovations-
2.html
10 - http://innovationkorea.eventbrite.com/

- Filtered Documents -----------------------------------
http://www.prweb.com/releases/2012/4/prweb9369229.htm
```

As we can see, the recommendations are all relevant and without duplications.

We executed the same query within 24 hours with the following results:

```
1 - http://jonjost.wordpress.com/2012/03/23/my-brilliant-academic-career-1/
2 - http://finance.yahoo.com/news/idiada-test-shows-zt1000-zeetex-
132100933.html
3 - http://finance.yahoo.com/news/whirlpool-corporation-comments-u-
department-171100217.html
4 - http://www.bbc.co.uk/news/uk-17508360
5 - http://www.huffingtonpost.com/muhammad-h-zaman/jim-yong-
kim_b_1377769.html
6 - http://adamsmithslostlegacy.blogspot.com/2012/03/james-otteson-on-
socialism-and-china.html
7 - http://freedomistheanswer.blogspot.com/2012/03/free-kibbles_25.html
8 - http://rogerpielkejr.blogspot.com/2012/03/manufacturing-and-industrial-
r.html
```

The resulting content was ordered by the associated measure of lexical entropy with the most important document on the top.

In the first set, the similarity subsystem was able to identify the similar articles that were tagged with the very high similarity value of 0.98. This was an indication that both the Article Extraction component and the Spatial Cosine Similarity provided a high quality output.

Looking at these URL we discover that the two documents (8)

**http://www.prweb.com/releases/2012/4/prweb9369229.htm** and (11)

**http://news.yahoo.com/pwn-technologies-wins-iwa-project-innovation-award-2012-080040615.html** represent the same article published under two different news sites, but only the one published by news.yahoo.com made the final list and its ranking of (4) is an indication that its content is quite rich in information and diverse in language.

From a quick look at the ranking used in the resulting documents and we can verify that the Lexical Entropy measure is performing quite well, by putting content-rich documents at the very top in both sets obtained in the two different runs.

## 3.2 Competitive Intelligence for the new drug Pradaxa

After the launch of our new drug Pradaxa for atrial fibrillation two new competitive drugs Xarelto and Apixaban, requested FDA approval for similar indications. In order to discover any article regarding their approval in relation to our product Pradaxa, we used the discovery system to monitor the news media for the entire month of April and part of May.

The system was configured with the following code:

```
1. Query query = new Query("pradaxa and (xarelto or apixaban)");
2. WebDiscoverer discoverer = new WebDiscoverer();
3. discoverer.setMaxDocuments(10);
4. discoverer.discoverContent(query);
```

The query in this case was more complex than the previous example and relates the term pradaxa to either xarelto or spixaban (or both).

The initial run generated the following output:

```
Querying: BingNews
Retrieved 8 documents.

Querying: YahooNews
Retrieved 4 documents.

Querying phase for "pradaxa and (xarelto or apixaban)" completed.

SPATIAL COSINE SIMILARITY TABLE ----------
1 - http://www.businessweek.com/news/2011-11-28/pfizer-bristol-myers-eliquis-recommended-by-u-k-cost-agency.
2 - http://uk.reuters.com/article/2011/11/13/us-heart-bloodthinners-idUKTRE7AC0VP20111113
3 - http://www.inpharm.com/news/170089/nice-okays-eliquis-blood-clots-after-surgery
4 - http://www.inpharm.com/news/169731/xarelto-gains-lucrative-new-licence-fda
5 - http://www.medpagetoday.com/Cardiology/Strokes/29411
6 - http://abcnews.go.com/Health/WellnessNews/fda-okays-xarelto-stroke-prevention/story?id=14882668
7 - http://www.marketwatch.com/story/frost-sullivan-large-atrial-fibrillation-population-with-unmet-needs-triggers-growth-in-the-us-anticoagulant-market-2012-04-03
```

```
8 - http://www.medpagetoday.com/Cardiology/VenousThrombosis/31044
9 - http://in.reuters.com/article/2012/01/09/us-drugs-pradaxa-
idINTRE8082HD20120109
10 - http://in.reuters.com/article/2012/01/09/us-drugs-pradaxa-
idINTRE8082HD20120109
11 - http://www.wallstreet-online.de/nachricht/4892295-j-p-morgan-cazenove-
stuft-bayer-neutral
12 - http://www.thepharmaletter.com/file/109731/eu-backs-bayers-xarelto-for-
stroke-prevention.html

1 - *1.00  0.35  0.45  0.36  0.28  0.31  0.16  0.26  0.34  0.34  0.16  0.23
2 -  0.00 *1.00  0.30  0.39  0.28  0.25  0.18  0.36  0.41  0.41  0.31  0.24
3 -  0.00  0.00 *1.00  0.33  0.26  0.33  0.21  0.27  0.30  0.30  0.22  0.20
4 -  0.00  0.00  0.00 *1.00  0.29  0.40  0.20  0.33  0.37  0.37  0.22  0.30
5 -  0.00  0.00  0.00  0.00 *1.00  0.30  0.17  0.30  0.32  0.32  0.22  0.15
6 -  0.00  0.00  0.00  0.00  0.00 *1.00  0.17  0.28  0.28  0.29  0.31  0.27
7 -  0.00  0.00  0.00  0.00  0.00  0.00 *1.00  0.19  0.12  0.12  0.01  0.26
8 -  0.00  0.00  0.00  0.00  0.00  0.00  0.00 *1.00  0.33  0.33  0.19  0.17
9 -  0.00  0.00  0.00  0.00  0.00  0.00  0.00  0.00 *1.00 *1.00  0.16  0.18
10-  0.00  0.00  0.00  0.00  0.00  0.00  0.00  0.00  0.00 *1.00  0.16  0.18
11-  0.00  0.00  0.00  0.00  0.00  0.00  0.00  0.00  0.00  0.00 *1.00  0.17
12-  0.00  0.00  0.00  0.00  0.00  0.00  0.00  0.00  0.00  0.00  0.00 *1.00
Calculating Document Similarities on 12 documents
Calculating Document Similarities on 11 documents
Calculating Document Similarities on 10 documents
The corpus size after filtering is of 10 documents.

- Selected Documents -----------------------------------
1 - http://www.marketwatch.com/story/frost-sullivan-large-atrial-
fibrillation-population-with-unmet-needs-triggers-growth-in-the-us-
anticoagulant-market-2012-04-03
2 - http://www.medpagetoday.com/Cardiology/VenousThrombosis/31044
3 - http://www.inpharm.com/news/170089/nice-okays-eliquis-blood-clots-after-
surgery
4 - http://www.inpharm.com/news/169731/xarelto-gains-lucrative-new-licence-
fda
5 - http://uk.reuters.com/article/2011/11/13/us-heart-bloodthinners-
idUKTRE7AC0VP20111113
6 - http://abcnews.go.com/Health/WellnessNews/fda-okays-xarelto-stroke-
prevention/story?id=14882668
7 - http://in.reuters.com/article/2012/01/09/us-drugs-pradaxa-
idINTRE8082HD20120109
8 - http://www.medpagetoday.com/Cardiology/Strokes/29411
9 - http://www.businessweek.com/news/2011-11-28/pfizer-bristol-myers-eliquis-
recommended-by-u-k-cost-agency.html
10 - http://www.thepharmaletter.com/file/109731/eu-backs-bayers-xarelto-for-
stroke-prevention.html

- Filtered Documents -----------------------------------
http://in.reuters.com/article/2012/01/09/us-drugs-pradaxa-
idINTRE8082HD20120109
http://www.wallstreet-online.de/nachricht/4892295-j-p-morgan-cazenove-stuft-
bayer-neutral
```

As we can see, the system detected the presence of and removed duplicate articles and also filtered the document "`http://www.wallstreet-online.de/nachricht/4892295-j-p-morgan-cazenove-stuft-bayer-neutral`" which was not in English.

From the result we can see how the lexical entropy measure provides a very good ordering of documents: from the very rich and informative article at the top of the list to the very small content represented by the bottom URL.

In the following days the WTDS kept monitoring the Internet, providing alerts as new content kept flowing such as the case of a breaking news story on May 23th regarding the rejection of Rivaroxaban (another name for Xarelto) by the FDA for the treatment of acute coronary syndrome. This news was immediately communicated to the marketing team and to upper management as it represented an important breakthrough for the marketing of Pradaxa and demonstrated the usefulness of this discovery platform.

# 4. Conclusions

Unstructured content published over the Internet represent an important, and often critical, source of information for any company that carries out competitive intelligence analysis. Yet, the enormous volume of data being generated every day from an ever-increasing number of sources makes it impossible for a single person, and even for a well-staffed team, to capture all the insight contained in such a corpus of documents.

The use of internal search engines, a practice that is increasing year by year, provides only a partial solution to this problem since web crawlers have inherent limitations in terms of scalability and visibility on internet content. Therefore, Business Intelligence groups are left with the unsatisfied need of automatic methods that are able to provide to systems, internal to the enterprise, a list of URLs representing internet content related to latest events that might be highly important for competitive purposes.

To satisfy this need, they rely on external companies for the gathering of such content. This is costly, prone to human error, and might result in noisy data with the real potential of hindering their analytical efforts.

However, it is possible to leverage current search services that are publicly available, like Google, Bing and Yahoo, to cite the most recognizable ones  and to index new content very close to their moment of creation. This will allow the gathering of a large collection of web content that can be processed by applying Text Analytics techniques to automatically refine the collection to a set of documents that are highly relevant to any analytical effort.

We presented a processing flow in which the topic, represented by a search query, is used to enquiry publicly available search engines. The result is then aggregated and manipulated to extract the textual information of interest and enhanced through the use of linguistic approach like tokenization and stemming.

Duplication and redundancy is kept to a minimum by the removal of duplicated URLs and similar documents. In order to consider the structural similarity of documents we introduced a new measure of similarity: Spatial Cosine Similarity, which combines the efficacies of word frequencies and spatial features for a more stable and reliable measure of similarity.

Finally, lexical and entropy measures are used to rank the remaining documents in order to provide a tangible quantity representing the complexity and amount of information contained in the textual information.

The system will continuously scan the World Wide Web providing fresh and timely updates to the research topic guaranteeing results that are relevant and free of duplications in a timely fashion.

We also wanted to provide a "Plug & Play" feature to the system with an architecture that allows the easy development and integration of new sources of data in order to extend the content gathering capabilities of the platform. This would play an important role for specialty sources like Deep Web, semantic data, RSS feed, etc.

The system described in this paper is currently being evaluated within our company and could represent an important change on how the competitive intelligence group gathers external information with a potential of savings in the order of tens of thousands of dollars.

We already found instances in which the discovery platform was able to detect signals in the news media stream related to competitive products, as described in the "Competitive Intelligence for the new drug Pradaxa" use case presented in the previous section, allowing the competitive intelligence community to be promptly alerted and act on important news as soon as it was released to the public.

The current approach can easily be extended with classification capabilities, semantic analysis and so forth in order to build relationships among the various content identified by the system for a more in-depth discovery of patterns in the identified topics.

# 5. Appendix

## 5.1. Determining the number of reduction steps

The following are the complete results of the 10 tests conducted on extracting the article on 10 random samples of web pages using different values for the reduction step parameter *k*.

**Sample 1**

| Article # | K=3 | K=4 | K=5 | K=6 | K=7 | K=8 | K=9 |
|---|---|---|---|---|---|---|---|
| 1 | 3 | 3 | 3 | 3 | 3 | 3 | 3 |
| 2 | 3 | 3 | 3 | 3 | 3 | 3 | 3 |
| 3 | 3 | 5 | 5 | 5 | 5 | 5 | 5 |
| 4 | 0 | 0 | 0 | 0 | 0 | 0 | 0 |
| 5 | 3 | 3 | 5 | 4 | 4 | 4 | 4 |
| 6 | 5 | 5 | 5 | 3 | 3 | 3 | 3 |
| 7 | 3 | 3 | 4 | 4 | 4 | 4 | 4 |
| 8 | 2 | 2 | 2 | 2 | 5 | 5 | 5 |
| 9 | 2 | 1 | 2 | 2 | 5 | 5 | 5 |
| 10 | 3 | 3 | 3 | 3 | 3 | 3 | 3 |

1. http://www.co.scurry.tx.us/ips/cms/
2. http://yugioh.wikia.com/wiki/Appropriate
3. http://stress.about.com/od/stressmanagementglossary/g/psychosomatic.htm
4. http://inspirationfeed.com/inspiration/websites-inspiration/45-outstandingly-well-designed-e-commerce-websites/
5. http://www.d20srd.org/srd/spells/mislead.htm
6. http://www.tantamount.com/words/
7. http://www.jbhifi.com.au/jb-hi-fi-home-audio/
8. http://tuft.ca/technology/
9. http://disneyland.disney.go.com/plan/guest-services/lost-and-found/
10. http://www.ebay.com/sch/i.html?_nkw=ovation

**Sample 2**

| Article # | K=3 | K=4 | K=5 | K=6 | K=7 | K=8 | K=9 |
|---|---|---|---|---|---|---|---|
| 1 | 1 | 3 | 3 | 4 | 5 | 5 | 5 |
| 2 | 5 | 5 | 5 | 5 | 5 | 5 | 5 |
| 3 | 2 | 2 | 1 | 2 | 5 | 5 | 5 |
| 4 | 1 | 1 | 2 | 2 | 2 | 2 | 2 |
| 5 | 5 | 5 | 5 | 5 | 5 | 5 | 5 |
| 6 | 5 | 5 | 5 | 5 | 5 | 5 | 5 |
| 7 | 5 | 5 | 5 | 5 | 5 | 5 | 5 |
| 8 | 3 | 3 | 3 | 3 | 3 | 3 | 5 |
| 9 | 5 | 5 | 5 | 5 | 5 | 5 | 5 |
| 10 | 2 | 2 | 2 | 3 | 3 | 4 | 4 |

1. http://www.salon.com/2011/05/04/torture_48/
2. http://www.subpop.com/artists/pissed_jeans
3. http://thesouthern.com/news/local/education/92f6e764-5989-11e1-bd34-001871e3ce6c.html

4. http://allrecipes.com/Recipes/Meat-and-Poultry/Chicken/Fried/
5. http://www.jewishjournal.com/picks_clicks/article/calendar_picks_and_clicks_february_1824_20120216/
6. http://www.techterms.com/definition/flaming
7. http://www.lining.com/EN/home/index.html
8. http://www.sibelius.com/products/scorch/index.html
9. http://www.opendoorsusa.org/persecution/
10. http://www.brendanbrazier.com/bio/index.html

**Sample 3**

| Article # | $K$=3 | $K$=4 | $K$=5 | $K$=6 | $K$=7 | $K$=8 | $K$=9 |
|---|---|---|---|---|---|---|---|
| 1 | 5 | 5 | 4 | 4 | 4 | 4 | 4 |
| 2 | 2 | 2 | 3 | 3 | 3 | 4 | 5 |
| 3 | 4 | 4 | 3 | 3 | 3 | 3 | 3 |
| 4 | 2 | 2 | 2 | 2 | 2 | 2 | 2 |
| 5 | 5 | 5 | 5 | 5 | 5 | 5 | 5 |
| 6 | 3 | 3 | 3 | 5 | 5 | 5 | 5 |
| 7 | 2 | 2 | 2 | 2 | 2 | 2 | 2 |
| 8 | 2 | 2 | 2 | 2 | 2 | 2 | 2 |
| 9 | 2 | 2 | 1 | 1 | 1 | 1 | 1 |
| 10 | 3 | 2 | 5 | 5 | 5 | 5 | 5 |

1. http://www.morewords.com/word/luxuriously/
2. http://www.absoluteastronomy.com/topics/Flash_flood
3. http://www.town.rye.nh.us/pages/index
4. http://www.mayoclinic.com/health/monoclonal-gammopathy/DS00870
5. http://www.wordiq.com/definition/Logically_possible
6. http://www.hydro-ofs.com/irrigate/
7. http://www.staples.com/Magnetic-Tape/product_SS1005128
8. http://download.oracle.com/javase/1.4.2/docs/api/java/awt/Component.html
9. http://tabs.ultimate-guitar.com/t/tristan_prettyman/madly_crd.htm
10. http://www.medicare.gov/navigation/medicare-basics/medicare-benefits/part-a.aspx

**Sample 4**

| Article # | K=3 | K=4 | K=5 | K=6 | K=7 | K=8 | K=9 |
|---|---|---|---|---|---|---|---|
| 1 | 5 | 5 | 5 | 5 | 5 | 5 | 5 |
| 2 | 3 | 3 | 3 | 3 | 3 | 3 | 3 |
| 3 | 2 | 2 | 3 | 4 | 5 | 2 | 2 |
| 4 | 2 | 2 | 2 | 2 | 2 | 3 | 3 |
| 5 | 2 | 2 | 3 | 3 | 3 | 3 | 3 |
| 6 | 2 | 2 | 3 | 3 | 3 | 3 | 3 |
| 7 | 3 | 3 | 4 | 4 | 4 | 5 | 5 |
| 8 | 3 | 3 | 3 | 3 | 3 | 3 | 3 |
| 9 | 2 | 2 | 2 | 2 | 2 | 2 | 2 |
| 10 | 5 | 4 | 5 | 5 | 5 | 5 | 5 |

http://www.redorbit.com/news/technology/1112476299/fcc-hangs-up-on-robocalls/
http://pediatrics.aappublications.org/content/current
http://www.absoluteastronomy.com/topics/Midsummer
http://music.yahoo.com/unkind/
http://informality.wordpress.com/category/definitions/
http://www.fifa.com/worldranking/rankingtable/index.html
http://www.masc.mb.ca/masc.nsf/index.html
http://www.flutebar.com/en/
http://www.nme.com/nme-video/youtube/id/HvI9VIHggRw/search/overstep
http://www.beaumontenterprise.com/business/press-releases/article/CherryFusion-Names-Dave-Wares-President-3332265.php

**Sample 5**

| Article # | K=3 | K=4 | K=5 | K=6 | K=7 | K=8 | K=9 |
|---|---|---|---|---|---|---|---|
| 1 | 2 | 2 | 3 | 3 | 3 | 5 | 5 |
| 2 | 5 | 5 | 5 | 5 | 5 | 5 | 5 |
| 3 | 2 | 2 | 5 | 5 | 5 | 5 | 5 |
| 4 | 5 | 5 | 5 | 5 | 5 | 5 | 5 |
| 5 | 3 | 3 | 3 | 3 | 3 | 3 | 3 |
| 6 | 3 | 5 | 5 | 5 | 5 | 5 | 5 |
| 7 | 2 | 2 | 3 | 3 | 3 | 3 | 5 |
| 8 | 1 | 1 | 1 | 1 | 1 | 1 | 1 |
| 9 | 5 | 5 | 5 | 5 | 5 | 5 | 5 |
| 10 | 1 | 1 | 2 | 3 | 5 | 5 | 5 |

1. http://tearsbeforepearls.blogspot.com/2012/01/it-has-been-busy-week-with-building.html
2. http://theinternetbachelorette.wordpress.com/2012/02/13/centerpiece-accessory/
3. http://www.businessweek.com/news/2012-02-17/singapore-lifts-aid-for-poor-as-budget-curbs-foreign-workers.html
4. http://democracyforburma.wordpress.com/2012/02/03/mahn-nyein-maung-top-karen-leader-facing-treason-charge/
5. http://themorningsun.com/articles/2012/02/18/sports/srv0000020952665.txt
6. http://www.redstate.com/natek58/

7. http://chicagotribune.feedsportal.com/c/34253/f/622872/s/1cb630a7/l/0L0Schicagotribune0N0Csports0Cbreaking0Cct0Espt0E0A2170Eautos0Enascar0E0E20A120A2170H0A0H2332790A0Bstory0Dtrack0Frss/story01.htm
8. http://austinlivetheatre.blogspot.com/2012/02/arts-reporting-terry-teachout-reports.html
9. http://w4.stern.nyu.edu/volatility/
10. http://markreckons.blogspot.com/2012/02/proof-government-prefers-dogma-on-drugs.html

**Sample 6**

| Article # | K=3 | K=4 | K=5 | K=6 | K=7 | K=8 | K=9 |
|---|---|---|---|---|---|---|---|
| 1 | 2 | 2 | 3 | 3 | 3 | 3 | 3 |
| 2 | 4 | 4 | 4 | 4 | 4 | 5 | 5 |
| 3 | 2 | 2 | 3 | 3 | 4 | 4 | 4 |
| 4 | 2 | 2 | 3 | 3 | 3 | 3 | 5 |
| 5 | 1 | 1 | 5 | 5 | 5 | 5 | 5 |
| 6 | 5 | 5 | 5 | 1 | 1 | 1 | 1 |
| 7 | 1 | 1 | 5 | 5 | 5 | 5 | 5 |
| 8 | 1 | 1 | 1 | 1 | 1 | 1 | 1 |
| 9 | 3 | 3 | 3 | 3 | 5 | 5 | 5 |
| 10 | 5 | 5 | 1 | 1 | 1 | 1 | 1 |

1. http://www.wpix.com/news/wpix-subway-slasher-gets-25-more-years,0,6474656.story?track=rss
2. http://www.wwlp.com/dpp/news/local/hampden/can-the-almanac-predict-the-weather
3. http://sftandthe101challenge.blogspot.com/2012/01/day-274-your-hair-do-you-scrimp-or-save.html
4. http://www.businessweek.com/news/2011-11-30/wall-street-unoccupied-as-200-000-job-cuts-bring-darkest-days-.html
5. http://theviewfromcullingworth.blogspot.com/2011/10/you-might-as-well-try-to-train.html
6. http://macdailynews.wordpress.com/2012/02/13/apples-ibooks-author-lets-mac-users-build-astounding-texts/
7. http://involvedcitizenry.blogspot.com/2011/08/city-of-brooklyn-park-residents-almost.html
8. http://www.autospies.com/news/TRUE-or-FALSE-Is-The-Fisker-Karma-The-World-s-Most-Interesting-Car-69229/
9. http://www.mtv.com/news/articles/1679474/whitney-houston-funeral-new-jersey.jhtml
10. http://rutlandherald.typepad.com/vermonttoday/2012/02/police-investigating-discovery-of-human-remains-in-danby.html

**Sample 7**

| Article # | K=3 | K=4 | K=5 | K=6 | K=7 | K=8 | K=9 |
|---|---|---|---|---|---|---|---|
| 1 | 2 | 2 | 4 | 4 | 4 | 4 | 5 |
| 2 | 2 | 2 | 5 | 5 | 5 | 5 | 5 |
| 3 | 3 | 3 | 3 | 3 | 5 | 5 | 5 |
| 4 | 2 | 2 | 3 | 3 | 5 | 5 | 5 |
| 5 | 1 | 2 | 4 | 5 | 5 | 5 | 5 |
| 6 | 2 | 2 | 5 | 5 | 5 | 5 | 5 |
| 7 | 2 | 3 | 3 | 3 | 4 | 4 | 4 |
| 8 | 2 | 2 | 2 | 2 | 3 | 3 | 3 |
| 9 | 3 | 3 | 3 | 4 | 4 | 5 | 5 |
| 10 | 3 | 3 | 3 | 3 | 3 | 3 | 4 |

1. http://msdn.microsoft.com/en-us/library/system.windows.forms.control.invalidate.aspx
2. http://www.nola.com/news/index.ssf/2012/02/metro_new_orleans_area_communi_211.html
3. http://alanlucker.blogspot.com/2011/05/28th-may-acent-of-bony-de-la-pica-and.html
4. http://offthedribble.blogs.nytimes.com/2012/02/14/30-seconds-with-paul-sturgess/?partner=rss&emc=rss
5. http://jasonvorhees.wordpress.com/2012/02/02/transformers-animated-oil-slick/
6. http://what-is-is.blogspot.com/2011/12/journalists-suddenly-disorderly-at-ows.html
7. http://societyofpoetry.wordpress.com/2011/11/13/ucla-what-do-do-if-you-are-a-winsome-writer-at-ucla-and-aint-got-no-writing-friends/
8. http://blastr.com/2012/02/david-hewlett-missing-mck.php
9. http://dearmissmermaid.blogspot.com/2011/04/klutzy-klutz-klutziness.html
10. http://newyork.cbslocal.com/2012/02/13/paul-tiny-sturgess-kicks-off-globetrotter-week-in-nyc/

**Sample 8**

| Article # | K=3 | K=4 | K=5 | K=6 | K=7 | K=8 | K=9 |
|---|---|---|---|---|---|---|---|
| 1 | 2 | 2 | 2 | 3 | 5 | 5 | 5 |
| 2 | 2 | 2 | 3 | 4 | 4 | 5 | 5 |
| 3 | 4 | 4 | 4 | 5 | 5 | 5 | 5 |
| 4 | 1 | 1 | 1 | 5 | 5 | 5 | 5 |
| 5 | 5 | 5 | 5 | 5 | 5 | 5 | 5 |
| 6 | 5 | 5 | 5 | 5 | 5 | 5 | 5 |
| 7 | 4 | 4 | 4 | 5 | 5 | 5 | 5 |
| 8 | 3 | 3 | 3 | 3 | 3 | 3 | 3 |
| 9 | 3 | 3 | 5 | 5 | 5 | 5 | 5 |
| 10 | 5 | 5 | 5 | 5 | 5 | 5 | 5 |

1. http://lacrossetribune.com/couleenews/lifestyles/june-dairy-days-gets-rolling-with-new-members/article_cbf6524a-58d3-11e1-bde7-001871e3ce6c.html
2. http://www.brooklyntabernacle.org/missions
3. http://the-adventurers-club.typepad.com/the_adventurers_club/2011/11/bangles-hazy-shade-of-winter.html
4. http://www.ibtimes.com/articles/300866/20120218/wills-lifestyle-india-fashion-week-2012-perfect.htm
5. http://www.joystiq.com/2012/02/16/civilization-5-gods-and-kings-expansion-announced-available-lat/
6. http://www.joystiq.com/2012/02/16/minecraft-lego-set-out-this-summer-check-out-the-first-pics/
7. http://www.yourhoustonnews.com/dayton/news/dayton-chamber-welcomes-timeless-treats/article_21ff8de8-377e-5a27-9b95-1d2564bdf77b.html
8. http://tnwordsmith.blogspot.com/2012/01/vote-for-riding-to-sundown-in-p-e-poll.html
9. http://www.businessweek.com/news/2012-02-18/ups-in-talks-to-buy-tnt-express-after-6-4-billion-bid-rejected.html
10. http://www.kcrg.com/news/local/Wellman-Couple-Remembered-as-Amazing-Parents-Leaders-139163259.html

Sample 9

| Article # | K=3 | K=4 | K=5 | K=6 | K=7 | K=8 | K=9 |
|---|---|---|---|---|---|---|---|
| 1 | 3 | 3 | 5 | 5 | 5 | 5 | 5 |
| 2 | 3 | 3 | 3 | 5 | 5 | 5 | 5 |
| 3 | 1 | 1 | 1 | 1 | 1 | 1 | 1 |
| 4 | 1 | 1 | 5 | 5 | 5 | 5 | 5 |
| 5 | 5 | 5 | 5 | 5 | 5 | 5 | 5 |
| 6 | 2 | 2 | 2 | 2 | 2 | 2 | 2 |
| 7 | 5 | 5 | 5 | 5 | 5 | 5 | 5 |
| 8 | 4 | 4 | 4 | 4 | 4 | 4 | 4 |
| 9 | 5 | 5 | 5 | 5 | 5 | 5 | 5 |
| 10 | 3 | 3 | 3 | 3 | 3 | 3 | 3 |

1. http://nausheenhusain.wordpress.com/2012/01/09/this-is-what-ive-learned-playing-ping-pong/
2. http://presbyterian.typepad.com/beyondordinary/2011/12/how-do-worshipers-decisions-about-church-giving-add-up-for-the-whole-church.html
3. http://www.summitdaily.com/article/20111219/NEWS/111219809/1011&parentprofile=1056ndtv.com
4. http://sewmuch2luv.blogspot.com/2011/11/davids-heptocopter-quilt.html
5. http://ralphcassar.wordpress.com/2011/10/14/suffocating/
6. http://www.investorwords.com/2337/homogeneous.html
7. http://www.eurekalert.org/pub_releases/2012-02/taac-ars021612.php
8. http://www.courant.com/community/county-hartford/hc-suffield-grief-recovery-0214-20120214,0,638886.story?track=rss
9. http://www.newsherald.com/news/avenue-100490-road-march.html
10. http://economicnerd.wordpress.com/2012/02/09/09-feb-2012-crude-silver/

Sample 10

| Article # | K=3 | K=4 | K=5 | K=6 | K=7 | K=8 | K=9 |
|---|---|---|---|---|---|---|---|
| 1 | 3 | 3 | 3 | 5 | 5 | 5 | 5 |
| 2 | 2 | 2 | 2 | 2 | 2 | 3 | 3 |
| 3 | 3 | 3 | 5 | 5 | 5 | 5 | 5 |
| 4 | 5 | 5 | 5 | 5 | 5 | 5 | 5 |
| 5 | 3 | 3 | 3 | 3 | 3 | 3 | 3 |
| 6 | 3 | 3 | 5 | 5 | 5 | 5 | 5 |
| 7 | 2 | 2 | 3 | 3 | 3 | 3 | 3 |
| 8 | 3 | 3 | 3 | 5 | 5 | 5 | 5 |
| 9 | 2 | 2 | 3 | 3 | 3 | 3 | 3 |
| 10 | 4 | 4 | 4 | 4 | 4 | 4 | 4 |

1. http://mindingmyownstitches.blogspot.com/2011/12/surmount-stash-2012.html
2. http://www.eoearth.org/article/Earthquake
3. http://finance.yahoo.com/news/endo-pharmaceuticals-present-j-p-213000509.html
4. http://www.thefader.com/2011/05/10/adeles-ex-boyfriend-wants-album-royalties-for-being-a-prick/
5. http://www.washingtonpost.com/sports/highschools/howard-county-girls-basketball-glenelg-edges-river-hill-to-clinch-regular-season-title-outright/2012/02/16/gIQAzOnpIR_story.html?wprss=rss_local
6. http://www.wthr.com/story/16959188/hearing-set-for-friday-in-bisard-case
7. http://figandsage.blogspot.com/2011/08/my-visit-about-face-body-natural-beauty.html
8. http://dfw.cbslocal.com/2012/02/15/cops-found-restraints-in-texas-torture-suspects-home/
9. http://www.nbcconnecticut.com/blogs/dog-house/WVU-Settles-With-Big-East-Headed-for-Big-12-139288858.html
10. http://biggovernment.com/awrhawkins/2012/02/17/dems-storm-out-of-hearing-on-government-overreach-because-it-was-a-hearing-on-government-overreach/email/

## 5.2 Determining the Similarity Threshold
*Set #1*

**Seed document 1**

*Sanctions choke off Iran oil output. Iran's oil production has fallen to a 10-year low and could drop to levels last seen during the Iran-Iraq war in the 1980s as sanctions over its nuclear programme disrupt an industry already suffering from years of underinvestment.*

**Seed document 2**

*Sanctions Reduce Iran oil output. Production of oil in Iran has fallen to a decade low and could drop to levels hardly seen during the last Iran-Iraq war in the 1980s as sanctions over its nuclear programme disrupt an industry already suffering from years of underinvestment.*

*Set #2*

*Seed document 1*:

*The U.S. could be compelled to sanction one of its closest Asian allies -- India -- unless the country makes significant progress by summer on cutting back on Iranian oil.*

*Under sanctions approved by Congress late last year, the United States is supposed to penalize any country that does business through Iran's central bank, which processes much of the country's oil transactions. Countries would be specifically penalized for oil purchases if they do not make a serious effort to curtail those Iranian imports.*

*Where exactly India stands in that effort is unclear. Earlier this year, Indian's finance minister said "it is not possible" to drastically reduce Iranian oil imports and that they're too vital to India's developing economy. Secretary of State Hillary Clinton, though, testified last month that India and other countries are doing more to comply "than perhaps their public statements might lead you to believe."*

*India technically is in the same boat as any other country when it comes to the U.S. sanctions. Come June, any country not in compliance could see their financial accounts in the U.S. shut down.*

*But India, the world's largest democracy, received additional attention this past week following a Bloomberg report that quoted unnamed officials saying India so far has not cut back on Iranian oil purchases -- and could be sanctioned.*

*Because of the country's fondness for Iranian oil, it is one of the most vulnerable to sanctions among U.S. allies.*

*According to data from the Energy Information Administration, the country is the third-biggest importer of Iranian crude oil -- averaging 328,000 barrels a day as of early 2011. China is the biggest importer, followed by Japan, another U.S. ally. South Korea and Turkey are also top importers.*

*A Treasury Department official told FoxNews.com on Friday that the administration is in the process of "talking to all countries" engaged in trade with Iran.*

*"We are working with our partners to significantly reduce their imports of Iranian crude and to isolate the Central Bank of Iran so as to limit the risks it can pose to the international financial system," the official said. "High level delegations from the Departments of the Treasury, State and Energy have already been traveling across the globe to consult with their counterparts on these issues."*

*Clinton was asked about those efforts -- particularly concerning India, China and Turkey -- during a hearing before the Senate Foreign Relations Committee last month. The secretary said the administration has held "very intense and very blunt conversations" with them on the issue, and suggested they could all avoid sanctions.*

*"I think that there are a number of steps that we are pointing out to them that we believe they can and should make," Clinton said. "I also can tell you that in a number of cases, both on their government side and on their business side, they are taking actions that go further and deeper than perhaps their public statements might lead you to believe. And we're going to continue to keep an absolute foot on the pedal in terms of our accelerated, aggressive outreach to them."*

*Bloomberg reported earlier in the week that India, despite public statements, is looking to start reducing Iranian imports in April.*

*Japan and South Korea also reportedly are looking at ways to reduce their imports, and avoid the sanctions.*

*President Obama since taking office has aggressively tried to build ties with Asian allies, at a time when China is rising. The administration's relationship with India has been far more trouble-free than its alliance with India's chief adversary, Pakistan, and the administration surely wants to avoid slapping on sanctions if possible.*

*Not only is the U.S.-India relationship important in terms of global politics, the business relationship between the two nations is a critical one. Late last year, Maryland Gov. Martin O'Malley and dozens other officials went on a trade mission to India, resulting in*

*millions of dollars in new contracts signed between the two countries. The Federation of Indian Chambers of Commerce and Industry also plans to open a joint center in Maryland, though the location has not yet been chosen.*

*Sanctions at this stage would be awkward, considering Obama tried to invigorate a new wave of U.S.-India relations when he visited the country in 2010.*

*In an implicit rebuke to China, Obama used that visit to back India's bid for a permanent seat on the U.N. Security Council.*

**Seed document 2**:

*India has failed to reduce its purchases of Iranian oil, and if it doesn't do so, President Barack Obama may be forced to impose sanctions on one of Asia's most important nations, Obama administration officials said yesterday.*

*A decision to levy penalties under a new U.S. law restricting payments for Iranian oil could come as early as June 28, according to several U.S. officials who spoke on condition of anonymity because of the sensitivity of the issue.*

*March 15 (Bloomberg) -- Alejandro Barbajosa, an oil markets specialist and business development manager at Argus Media Inc., talks about oil supply disruption and the role of the International Energy Agency. He speaks from Singapore with Caroline Hyde on Bloomberg Television's "First Look." (Source: Bloomberg)*

*"Given the level of trade, and in particular oil, between Iran and India, targeting an Indian entity that facilitates Iran's access to the international financial market should be top of mind for the U.S. Treasury," Avi Jorisch, a former Treasury Department official who is now a Washington-based consultant on deterring illicit finance, said in an interview.*

*The U.S. law, which targets oil payments made through Iran's central bank, applies to any country that doesn't make a "significant" reduction in its Iranian crude oil purchases during the first half of this year. If India fails to cut Iranian imports*

*sufficiently, Obama may be compelled to bar access to the U.S. banking system for any Indian bank processing oil payments through Iran's central bank, the U.S. officials said.*

*While India hasn't asked its refiners to stop purchasing Iranian crude, the government has told processors in the South Asian nation to seek alternate supplies and gradually reduce their dependence on the Persian Gulf state due to increasing pressure from the U.S. in recent weeks, three Indian officials with direct knowledge of the situation said today.*

*No Significant Reduction*

*India hasn't significantly cut imports this year because refiners' annual crude term deals with Iran typically run from April to March, they said. The planned reductions will start only when new annual contracts begin next month, the Indian officials said, declining to be identified because they aren't authorized to speak to the media.*

*India bought an average of 328,000 barrels a day of Iranian crude in the first six months of last year, making it the No. 3 buyer, behind China and Japan and ahead of South Korea, according to the U.S. Energy Information Administration. Iran is the No. 2 producer in the Organization of Petroleum Exporting Countries.*

*The U.S. government may not be aware that India's biggest buyer of Iranian oil, state-owned Mangalore Refinery & Petrochemicals Ltd. (MRPL), plans to import less from Iran starting next month, according to two officials with direct knowledge of the matter who spoke on condition of anonymity because they weren't authorized to speak.*

*Unilateral Sanctions*

*Oil Minister S. Jaipal Reddy, Finance Minister Pranab Mukherjee and Foreign Secretary Ranjan Mathai have said India will continue to buy Iranian oil to meet its growing energy needs. While the Indian government has an excellent record of enforcing United Nations sanctions on Iran, India has objected to unilateral U.S. sanctions, according to U.S. officials.*

*"We abide scrupulously by UN authorized sanctions," Indian Foreign Ministry spokesman Syed Akbaruddin said in a telephone interview. While restrictions imposed by individual countries "have an impact on commercial interactions, from a legal perspective there is nothing that binds us to follow them."*

*Oil Purchases Rise*

*The latest shipping data shows India and South Korea sharply increased oil purchases from Iran in January, according to a report released yesterday by the International Energy Agency in Paris. China halved its imports from Iran, from 550,000 barrels a day in December to 275,000 barrels a day in January, following a dispute over pricing terms that has now been resolved, the report said.*

*The new U.S. law targeting Iranian petroleum transactions doesn't specify by what percentage a nation must reduce its Iranian oil imports to qualify for an exemption from sanctions. U.S. officials, speaking on condition of anonymity, said they are looking for cuts of around 15 percent in volume, though they might consider whether buyers have extracted significant price discounts, thereby depriving Iran of revenue.*

*Mangalore Refinery may cut its contract to 6 million metric tons, or 120,000 barrels a day, in the year ending March 2013, which would be a 15 percent cut from the previous year, one of the people with knowledge of the planned reductions said.*

*U.S. Offer*

*The U.S. has offered India help in brokering deals with alternative suppliers including Iraq and Saudi Arabia, which has offered to replace any shortfall, according to U.S. and Indian officials.*

*Envoys from the White House, the State Department, the Treasury Department and the U.S. Embassy in India have had numerous conversations with Indian counterparts since Congress began debating the sanctions measure that Obama signed into law Dec. 31.*

*Nancy Powell, the Obama administration's ambassador- designate to India, testified before Congress last month that if confirmed, she would be "spending a great deal of*

*time" working with India on Iran sanctions issues. She quoted Mathai, who came to Washington for meetings last month, as saying India is working to diversify its sources of petroleum and reduce its dependence on Iran to no more than 10 percent of its total oil imports.*

*"That would be a very good sign," Powell said.*

*Disrupting Shipments*

*U.S. and European Union sanctions are already disrupting Iranian crude shipments to global refiners, contributing to a 16 percent advance in London-traded Brent this year. Brent oil for April settlement fell $1.25, or 1 percent, to end the session at $124.97 yesterday on the London-based ICE Futures Europe exchange.*

*The EU decided two months ago to embargo Iranian oil imports effective July 1. Last year, the 27-member EU was collectively the No. 2 importer of Iranian oil, taking 18 percent of Iran's crude exports. Faced with a shrinking pool of buyers, Iran last month offered India additional crude supplies on revised terms.*

*China's Ministry of Foreign Affairs has criticized sanctions on Iranian oil. China's crude imports from Iran hit their lowest level in five months in January, customs data show, as the country's biggest buyer, China International United Petroleum & Chemical Co. (0119173D), known as Unipec, delayed signing a new contract because of a dispute over payment terms. Unipec cut its 2012 term contract purchases by 15 percent from 2011, though the payment dispute has since been resolved.*

*Japan, Korea*

*For their part, Japan and South Korea are seeking exemptions from the new U.S. sanctions. If both nations can demonstrate a significant reduction in their purchases by the end of June, their banks would escape penalties, according to two U.S. officials involved in the talks.*

*Japan is seeking to reduce its crude purchases from Iran by at least 11 percent, according to a Japanese government official interviewed Feb. 21. The three largest*

*Japanese buyers of Iranian crude are Showa Shell Sekiyu KK (5002), JX Nippon Oil & Energy Corp. and Cosmo Oil Co. (5007)*

*The South Korean government has said it will make a decision on cutting Iranian crude imports by the end of June. South Korean officials denied reports saying they had already proposed cutting imports by 15 percent to 20 percent.*

*The White House doesn't want to punish Japan, South Korea or India, critical U.S. partners in trade and security and important regional counterweights to the rise of China, U.S. officials said. Still, the president has limited leeway to grant exemptions under the law, and so far, India hasn't demonstrated reductions, they said.*

*Free Pass*

*Mark Dubowitz, the executive director of the Foundation for the Defense of Democracies in Washington and an adviser to the administration on sanctions, said India shouldn't assume it will avoid sanctions unless its refiners demonstrably reduce imports over the next three months.*

*There's no reason "why India should be given a free pass as the EU, Japan and others significantly reduce the scale and scope of their Iranian trade," he said in an interview. "No country should be confident that it will not be the target of U.S. sanctions."*

*Other analysts said Indian officials have responded to U.S. pressure by quietly pressing state-run refiners to switch to alternative sources, and they expect the U.S. to reach an accommodation with the world's most populous democracy.*

*"It's highly unlikely that the U.S. would sanction India on this issue. The Iran issue is an irritant at best," Harsh V. Pant, a specialist on India and Iran at King's College London, said in a telephone interview.*

*The Iranian central bank sanctions that Obama signed into law Dec. 31 are part of a larger effort to deprive the Persian Gulf country of its leading source of revenue and complicate Iran's commercial ties with the outside world. The U.S. and the EU have piled on new sanctions since November in an effort to pressure Iran to abandon any work*

*it may be conducting to acquire a nuclear weapons capability. Iran insists that its nuclear program is strictly for civilian energy and medical research."*

## 5.3 Code

For a copy of the WTDS don't hesitate to contact me at giancarlo.crocetti@my.ccsu.edu.

# 6. Bibliography

Ahmadullin I, Damera-Venkata N., Fan J., Lee S., Lin Q, Liu J., & O'Brien-Strain E. (2011). *Document Visual Similarity Measure for Document Search*. ACM

Ananth Krishnan K. (2011, April 27). *The "Unstructured Information" Most Business Miss Out On*. The New Intelligent Enterprise

Arampatzis A., Van der Weide Th.P, Koster C.H.A., Van Bommel P. (1999). *An Evaluation of Linguistically-motivated Indexing Schemes.* Dept. of Information Systems and Information Retrieval, University of Nijmegen, The Netherlands.

Aslam J., & Frost M. (2003). *An Information-theoretic Measure for Document Similarity*. ACM

Bai X., Cambazoglu B., & Junquiera F. P. (2011). *Discovering URLs Through User Feedback* CIKM '11**:** Proceedings of the 20th ACM international conference on Information and knowledge management

Basu S. (Mar. 14th, 2010) *10 Search Engine to Explore the Invision Web*. Retrieved from http://www.makeuseof.com/tag/10-search-engines-explore-deep-invisible-web/

Carterette B., & Allan J. *Semiautomatic Evaluation of Retrieval Systems Using Document Similarities*. Center for Intelligence Information Retrieval, Dept. of Computer Science,

University of Massachusetts.

Dasgupta A., Kumar G. R., Olston C., Pandey S., & Tomkins A. (2007). *The Discoverability of the Web*. WWW '07: Proceedings of the 16th International Conference on World Wide Web, pages 421–430, New York, NY, USA. ACM

Duran P., Malvern D. & Richards B. (2004). *Developmental Trends in Lexical Diversity*. Oxford University Press.

Farkas J. *Using Kohonen Maps to Determine Document Similarity*. Centre for Information Technology Innovation (CITI), Quebec.

Feldman R., Sanger J. (2009). *The Text Mining Handbook, Advanced Approaches in Analyzing Unstructured Data.* Cambridge University Press

Goday R. (2011, November 23). *Search and Discovery: How They Complement Each Other*. Retrieved from http://www.darwineco.com/blog/bid/74792/Search-and-Discovery-How-They-Complement-Each-Other

Goday R. (2011, October 23). *6 Traits of Highly Effective Content Discovery Engines*. Retrieved from http://www.darwineco.com/blog/bid/70094/6-Traits-of-Highly-Effective-Content-Discovery-Engines

Golovin N., & Rahm E. (2011). Automatic Optimization of Web Recommendations Using Feedback and Ontology Graphs. University of Leipzig, Germany.

Kozorovistky A.K., & Kurland O. (2011). *From "Identical" to "Similar": Fusing Retrieved List Based on Inter-Document Similarities*. Journal of Artificial Intelligence Research 41:267-296.

Kurland O. (2006). *Inter-document similarities, language models, and ad hoc information retrieval*. PhD dissertation, Cornell University.

Halvorsen K., Hart T., Islam-Frenoy D., Puopolo J., Wessel P., & Zeier C. (2008). *Book of Search.* Microsoft Press.

Helmer S. (2007). *Measuring the Structural Similarity of Semistructured Documents Using Entropy*. ACM

Hugh C. (2010). *Entropy and Divergence in a Modern Fiction Corpus*. Digital Humanities Conference

Islam A.,Inkpen D. (2008). *Semantic Text Similarity Using Corpus-Based Word Similarity and String Similarity.* University of Ottawa.

Lakkaraju P., Gauch S., & Speretta M. (2008). *Document Similarity Based on Concept Tree*

*Distance*. ACM.

Lee N., & Cho K. (2011). *A Computational Linguistic Approach to Reviewer's Comments and Their Writing Abilities*. SungKyunkwan University Seoul, South Korea

Nelson A. (2002, November 1). *Lexical Diversity*. Retrieved from http://www.craweblogs.com/commlog/archives/000533.html

Pasternack J., Roth D. (2009). *Extracting Article Text from the Web with Maximum Subsequence Segmentation.* University of Illinois at Urbana-Champain

Schonfeld U., & Shivakumar N. (2009). *Sitemaps: Above and Beyond the Crawl of Duty*. WWW 2009 in Madrid.

Sun Z., Errami M., Long T., Renard C., Choradia N., & Garner H. (2012). *Systematic Characterization of Text Similarity in Full Text Biomedical Publication.* PLoS ONE

Thor A., Golovin N., Rahm E. (2005). *Adaptive website recommendations with AWESOME*. The VLDB Journal 14(4): 357-372. Springer-Verlag

Torres D.F.M. (2002). *Entrophy Text Analyzer.* Lisbon University.

Tsai M., Chen H.*Some Similarity Computation Methods in Novelty Detection.* National Taiwan

University

Weiss S. M., Indurkhya N., Zhang T., Damerau F. J. (2005). *Text Mining, Predictive Methods for Analyzing Unstructured Information.* Springer

Zillman M.P. (2012, January) – "Deep Web Research 2012" - http://deepweb.us/